\documentclass[10pt]{article}
\usepackage{latexsym,graphicx}
\usepackage{cite}
\usepackage{amsmath}
\usepackage{amsfonts}
\usepackage{amssymb}
\usepackage[left=2.5cm,top=2.5cm,right=2.5cm,bottom=1.5cm]{geometry}
\usepackage{graphicx}
\usepackage{epstopdf}
\usepackage{float}
\DeclareGraphicsExtensions{.eps}
\begin{document}
\begin{center}
\large{\bf{ Role of a periodic varying deceleration parameter in Particle creation with higher dimensional FLRW Universe }} \\
\vspace{10mm}
\normalsize{Priyanka Garg$^1$, Anirudh Pradhan$^2$}\\
\vspace{5mm}
\normalsize{$^{1,2}$Department of Mathematics, Institute of Applied Sciences \& Humanities, GLA University,\\ 
Mathura-281 406, Uttar Pradesh, India \\
\vspace{5mm}
$^1$E-mail:pri.19aug@gmail.com \\
\vspace{2mm}
$^2$E-mail: pradhan.anirudh@gmail.com }\\
\end{center}
\vspace{10mm}
\begin{abstract}
The present search focus on the mechanism of gravitationally influenced particle creation (PC) in higher dimensional 
Friedmann-Lemaitre-Robertson-Walker(FLRW) cosmological models 
with cosmological constant (CC). The solution of the corresponding field equations is obtained by assuming a periodically varying 
deceleration parameter (PVDP) i.e. $q= m \cos kt - 1$ [Shen and Zhao, Chin. Phys. Lett., 31 (2014) 010401] which gives a scale factor 
$a(t) = a_0 \left[\tan \left(\frac{kt}{2}\right)\right]^\frac{1}{m}$, where $a_0$ is the scale factor at the current epoch. Here $k$ displays 
the PVDP periodicity and can be regarded as a parameter of cosmic frequency, $m$ is an enhancement element that increases the PVDP peak. 
Here, we investigated periodic variation behavior of few quantities such as the deceleration parameter $q$, the 
energy density $\rho$, PC rate $\psi$, the entropy $S$, the CC $\Lambda$, Newton’s gravitational constant 
$G$ and discuss their physical significance. We have also explored the density parameter, proper distance, angular distance, luminosity distance, 
apparent magnitude, age of the universe, and the look-back time with redshift $z$ and have observed the role of particle formation in-universe 
evolution in early and late times. The periodic nature of various physical parameters is also discussed which are supporting the recent observations.
\end{abstract}


\smallskip

\section{Introduction}
The phenomenon of universe expansion rate becomes an attention point for all cosmologists, astrophysicists, and astronomers. This fact 
has been confirmed by observations (Supernovae Ia, CMB, BAO, etc). In 1998 and following years, 
some surprising results have been obtained by several groups of astronomers \cite{ref1,ref2,ref3,ref4,ref5,ref6,ref7} to estimate the 
universe expansion rate. These groups have estimated the separations and predicted the accelerated expansion of the cosmos, which is 
probably going to continue forever in view of SN Ia observations. It is predicted by these observations that something is responsible 
for this expansion. As a solution, the cosmological constant befits the context in this regard. Also, on large scales, the universe is 
having flat geometry as predicted by the cosmic microwave background (CMB) \cite{ref8,ref9,ref10,ref11,ref12,ref13}. Since there is not 
sufficient matter in the universe, so to deliver this flatness, the same 'dark energy' may be a candidate. In addition, the impact of DE 
appears to fluctuate, with the expansion of the universe decreasing and increasing over the period of time \cite{ref14,ref15,ref16,ref17}. 
CC $\Lambda$ is the simplest candidate for Dark Energy (DE) yet it should be, to a great degree, modified to fulfill 
the present estimation of the dark energy.\\

In present time a dynamical cosmological term $\Lambda(t)$ has attracted the attention of researchers as it resolve the cosmological 
constant problem. There is substantial observational evidence to detect Einstein’s cosmological constant, or a part of the universe’s 
material content that changes gradually over time to behave like $\Lambda$. 
The birth of the universe was caused by an excited vacuum fluctuation that caused the Super-fresh to adopt an inflationary expansion.
The release of the vacuum's stored energy results in subsequent heating. The cosmological term, which is the measurement of empty space energy, 
generates a repulsive force against the gravitational attraction between galaxies. A repulsive force against the gravitational attraction between
the galaxies is created by the cosmological term, which is the measurement of empty space energy. If the cosmological parameter occurs, since mass 
and energy are identical, the energy it describes counts as mass. If the cosmological term is huge enough, it may result in inflation by its 
energy plus the matter in the universe. \\

In the relativistic cosmological models the study of particle creation has drawn by many authors
\cite{ref18,ref19,ref20,ref21,ref22,ref23,ref24,ref25,ref26,ref27,ref28,ref28a,ref28b,ref28c,ref28d,ref28e,ref28f,ref28g,ref28h,ref28i,ref28j,ref28k,ref28l}. Prigogine \cite{ref29,ref30} gave the first theoretical approach of 
particle creation. Schrodinger \cite{ref31} has discussed the possibility of particle creation production as a result of space time curvature. 
Generally in curved space time a unique vacuum state does not exist. So there is ambiguity in the physical perception and the description of 
particles becomes even more complicated \cite{ref32,ref33}. There are two general approaches to understand the physical 
concept of particle production: (i) the technique of adiabatic vacuum state \cite{ref18} and (ii) the technique of instantaneous Hamiltonian 
diagonalization \cite{ref34}. Singh et al. \cite{ref20} discussed statefinder diagnostic in particle creation. Recently, Dixit et al. \cite{ref35} 
have searched particle creation in FLRW higher dimensional universe with gravitational and cosmological constants.\\

Recently a lot of cosmologists and astrophysicists are there addressed the FRW models with particle creation problem \cite{ref36,ref37,ref38}. 
Zimdahl, Yuan Qiang, and their colleague \cite{ref39,ref40} studied the models of particle creation with SN 1a data and showed the result is 
consistent and the universe is in an accelerating phase. At the moment, \cite{ref41} is exploring a different type of matter formation. 
Many researchers \cite{ref42} have recently paid great attention to the cosmology of the production of `adiabatic’ particles that are 
gravitationally induced and explain the present accelerated expansion. The particle outputs of non-minimally coupled scalar fields of light 
will lead to an early accelerated universe \cite{ref43} due to the change in space-time geometry.  \\

The new advances in super-string theory and super gravitational theory have inspired physicists' interest in exploring the 
evolution of the universe in higher-dimensional space-times.  Kaluza \cite{ref45} and Klein \cite{ref46} suggested an 
eminent five-dimensional theory in which gravity and electromagnetism are combined with additional dimensions. Many cosmologist \cite{ref47,ref48}
have worked on higher dimensional theory. A marvelous review of the higher dimensional unified theory which has an excellent discussion about 
the cosmological and astrophysical implications of extra dimensions been presented by Overduin and Wesson \cite{ref49}. For describing the 
nature of gravitational constant and cosmological constant, Harko and Mak \cite{ref56} has been proposed a different type of theory which is 
related to matter creation.   \\

The purpose and objective of our paper are to study the output of particles and the generation of entropy in the higher-dimensional FLRW model 
with periodically varying deceleration parameter. Here, in this model we discuss particle creation and entropy generation which has a Big Rip 
singularity \cite{ref57,ref58}. Moreover, we investigate some cosmological quantities such as the energy density $(\rho)$, the PC 
rate $(\psi)$, the entropy $(S)$, the deceleration parameter $(q)$, etc which are dependent on time. These quantities demonstrate the behavior of 
periodic variation with singularity-I. For an oscillating cosmic model, a time-dependent PVDP has been discussed 
in \cite{ref59}. Cosmological oscillating models have been also discussed in literature \cite{ref59a,ref60,ref61}.\\

The paper has the following structure. The derivation of the field equations in FLRW is presented in Section $2$. We find the solution 
of field equations for FLRW space times in Section $3$.  We outline estimates of some other physical and kinematic parameters using PVDPP 
in section $4$, this section has two subsections: the model with particle creation and the model without particle creation. 
Interpretation of the derived results has been given in section $5$. Kinematic tests are discussed in Section $6$.  Finally, conclusion are summarized 
in section $7$.

\section {Explicit Field Equations in FLRW}
Consider the Friedman-Lemaitre-Robertson-Walker (FLRW) metric for a (d+2)-dimensional homogeneous, isotropic, and flat model
\begin{equation}
\label{1}
ds^{2}= dt^{2} - a^{2}(t) \left[ {dr^{2}} + r^{2}  d{x_{d}^2}  \right]
\end{equation}
where $a$ is time dependent scale factor, and $dx_d^{2}$ define as
\begin{equation}
\label{2}
dx^{2}_d = d \theta_{1}^2 + sin^2 \theta_{1}  d \theta_{2}^2 + ........
 sin^2 \theta_{d-1}  d \theta_{d}^2
\end{equation}
Considering that the universe is full of a perfect fluid whose energy-momentum tensor is defined as 
\begin{equation}
\label{3}
T_{ij} = \rho \left(1 + \frac{p}{\rho}\right){\mu_i}{\mu_j} - p{g_{ij}} .
\end{equation}
Here $\rho$ $\&$ $p$ stand for the energy density and pressure of the fluid respectively, $u_i$ is the (d + 2) velocity vector 
which satisfy $u_i u^i = 1$. Einstein field equations (EFEs) with time-varying $\Lambda$ are given by

\begin{equation}
\label{4}
R_{ij} - \frac{R}{2}  g_{ij} = - \Lambda g_{ij} - 8 \pi G T_{ij}, 
\end{equation}
where $R_{ij}$, $g_{ij}$, and $R$ are the Ricci tensor, the metric
tensor and the Ricci scalar, respectively. And $G$ $\&$ $\Lambda$ indicates time dependent gravitational constant and cosmological
constant. By Eqs. (\ref{1}) and (\ref{3}), the EFEs (\ref{4}) reduce to  
\begin{equation}
\label{5}
\frac{(d^2 + d)}{2} H^2 = 8\pi G \rho + \Lambda
\end{equation}
and 
\begin{equation}
\label{6}
d (\dot{H} + H^2) + \frac{(d^2-d)}{2} H^2 = -8\pi G p + \Lambda,
\end{equation}
where $H = {\left(\frac{\dot a}{a}\right)}$ and dot shows a derivative with respect to cosmic time ($t$). On differentiating Eq. (\ref{5}), we find
\begin{equation}
\label{7}
{(d^2 + d)}{H \dot H} = 8\pi  (\dot {G}(t) \rho + {G} {\dot{\rho}(t)}) + \dot{\Lambda}(t)
\end{equation}
Multiplying by $(-1)$ in Eq. (\ref{5}) and adding with Eq (\ref{6}), we find,
\begin{equation}
\label{8}
 d \dot{H} = - 8\pi G (\rho + p) 
\end{equation}
On solving Eqs. (\ref{7}) and (\ref{8}), continuity equation is obtained as
\begin{equation}
\label{9}
{\dot{\rho}} + \rho (d+1) \left(1 + \frac{p}{\rho}\right) H = - \left( \rho \frac{\dot G}{G} +
\frac{\dot{\Lambda} }{ 8\pi G} \right) 
\end{equation}
With the help of Eqs. (\ref{5}) and (\ref{9}), we obtain 
\begin{equation}
\label{10}
\rho (t) + p = - \frac{d \dot{H}}{8 \pi G},
\end{equation}
\begin{equation}
\label{11}
 \Lambda(t) = \frac{(d^2+d)}{2} H^{2} - 8\pi G {\rho},
\end{equation}
\section{Particle Development Thermodynamics}
We assume that the early universe particle substance consists of a non-interacting relativistic fluid having a particle number density $n$ and 
equation of state (EoS) as follows:

\begin{equation}
\label{12}
 \omega  = \frac{p}{\rho},
\end{equation}
and
\begin{equation}
\label{13}
n = n_{0} \left(\frac{\rho}{\rho_{0}}\right)^{(\frac{1}{1+ \omega})},
\end{equation}
where $\omega$ is EoS parameter which lies in the interval $-1 <\omega\le{1} $.
However, the Supernova SN 1a, and Cosmic background radiation
(CMB) data \cite{ref11} show that for the accelerating universe, equation of state parameter is lying in the range 
 $-1.3 <\omega\le{-0.79} $. $n_{0} \geq 0$  and $\rho_{0} \geq 0$ are the current values of the particle number and energy density
The particle number density gives the equilibrium equation
\begin{equation}
\label{14}
  \dot {n}+ n H (1+ d) = n \psi {(t)}.
\end{equation}
Here $\psi (t)$ stands for time-dependent PC rate. On the side, $\psi (t) > 0$ show a particle source, $\psi (t) < 0 $ indicate 
particle disappearance and $\psi (t) = 0$ gives no particle production.\\

Using Eqs. (\ref{12}), (\ref{9}) $\&$ (\ref{13}), we obtain
\begin{equation}
\label{15}
\frac{1}{\rho} \frac{d \rho}{dt} + (d+1) (\omega + 1) H = -  \left(\frac{\dot G}{G} + \frac{\dot{\Lambda} }{ 8\pi G \rho}\right) ,
\end{equation}

\begin{equation}
\label{16}
  \frac{\dot {n}}{n} = \frac{1}{(1+ \omega)\rho } \frac{d \rho}{dt}.
\end{equation}

To find particle creation function we use Eqs. (\ref{14}) $-$ (\ref{16}):
\begin{equation}
\label{17}
\psi(t) = -\left(\frac{1}{ 1+ \omega }\right)  \left({\frac{\dot G(t)}{G(t)}} + \frac{\dot{\Lambda} (t)}{ 8\pi \rho G(t)} \right) .
\end{equation}

The entropy $S$ generated during PC at temperature $T$ follows this relation

\begin{equation}
\label{18}
p \frac{dV}{dt} + \frac{d(\rho V)}{dt} = T \frac{dS}{dt},
\end{equation}
where $V = a^{(d+1)}$ is a spatial volume. In cosmological fluid, entropy is define as
\begin{equation}
\label{19}
S =  \frac{{\rho}(1+ \omega){a^{(d+1)}}}{T}.
\end{equation}
From Eq. (\ref{18}) and Eq. (\ref{9}) we find,
\begin{equation}
\label{20}
 \frac{dS}{dt} = \frac{(1+ \omega){\rho}{a^{(d+1)}}}{T} {\psi(t)}.
\end{equation}
The entropy as a function of $\psi(t)$ (PC rate) is obtained from the Eq. (\ref{20}) as
\begin{equation}
\label{21}
S(t) = S_{0} e^{\int {\psi(t)}dt},
\end{equation}
where $S_0$ is an integrating constant. Here, we have summarized the formation of PC $\psi(t)$ and entropy $S(t)$ which 
is time-dependent and based on the previous studies. In these references \cite{ref29,ref30,ref56,ref62}, we can found discussion in detail. 

\section{Role of PVDP in Solutions}
We consider a time-dependent periodic varying deceleration parameter (PVDP) \cite{ref59} to solve field equations as 

\begin{equation}
\label{22}
q= m \cos(k t)-1,
\end{equation}
where $k$ $\&$ $m$ figures as $+ve$ constants. Here the periodicity of the PVDP is defined by $k$ which can be viewed as a parameter of 
cosmic frequency. The peak of the PVDP is increased by the enhancement factor $m$. In this model, the universe begins with a period of 
deceleration and expands in a cyclic history into a phase of super-exponential growth.  Here by the definitions $q=-1-\frac{\dot H}{H^{2}}$ $\&$ 
$\dot{a}=a H$ we obtain $\dot{H}=-m H^{2}$ cos $k t$. \\

After integration of Eq. (\ref{22}), we obtain
\begin{equation}
\label{23}
H=\frac{k}{ k_{1} + m ~\sin(k t) }
\end{equation}

where $k_{1}$ is an integrating constant. We may consider $k_{1}=0$ without losing generality.
Hence the Hubble function becomes  
 
\begin{equation}
\label{24}
H=\frac{k}{m ~\sin (k t)}
\end{equation}

By integration of Eq. (\ref{24}), we find the $a(t)$ (scale factor) such as, 
\begin{equation}
\label{25}
a(t)= a_0 \left[[tan \left(\frac{kt}{2}\right)\right]^\frac{1}{m} .
\end{equation}
where $a_0$ is the scale factor at the present epoch. \\

By using the constraints $H_{0}=69.2$ and $q_{0}=-0.52$  from the recent observational Hubble data (OHD) and joint light curves (JLA) data 
\cite{ref63,ref64} in Eq. (22), we find a relation between $k$ $\&$ $m$ i.e. $k = H_0 ~ cos^{-1} \left( {\frac {q_0 + 1}{m}} \right)$. From 
this relation we obtain the value of $m$ for different value of $k$. \\

Now, in these subsections, we obtain the cosmological solutions of with particle creation $\&$ without particle creation.


\subsection{With Particle Creation}
Using equations (\ref{10})-(\ref{12}) $\&$ (\ref{24}), we find time dependent energy density $\rho$ $\&$ time dependent cosmological 
constant $\Lambda$, respectively as
 \begin{equation}
 \label{26}
\rho = {\frac {d{k}^{2}\cos \left( kt \right) }{8m \left( \sin \left( kt
		\right)  \right) ^{2} \left( \omega+1 \right) \pi \,G}}
 ~,
 \end{equation}
 
 \begin{equation}
 \label{27}
 \Lambda = {\frac {d \left( d+1 \right) {k}^{2}}{2{m}^{2} \left( \sin \left( 
 		kt \right)  \right) ^{2}}}-{\frac {d{k}^{2}\cos \left( kt \right) }{m
 		\left( \sin \left( kt \right)  \right) ^{2} \left( \omega+1 \right) }}.
\end{equation} 
 It is obviously that, to be valid for both solutions $\omega \neq -1$ is required. \\
 
 Equations (\ref{17}), (\ref{26}) and (\ref{27}) gives particle creation rate as

\begin{equation}
\label{28}
\psi =
 {\frac {  \left( d+1 \right) k}{m\sin
		\left( kt \right) }}+ \left( {\frac {{-2 k} \left( \cos
		\left( kt \right)  \right) }{  \sin \left( kt \right) 
		 }}- {\frac {{k \sin \left( kt \right) }}
	{\cos \left( kt \right)  }} \right) 
 \left( \omega+1 \right).
\end{equation}

Finally, we obtained the entropy production during the particle creation from Eqs. (\ref{21}) and (\ref{28}) as

\begin{equation}
\label{29}
S(t) = S_{0} exp \left[ {\frac {4 \ln  \left( \csc \left( kt
		\right) - (d+1) (\omega+1) \cot \left( kt \right)  \right) d}{ \left( \omega+1 \right) 
		m}} + {\frac {  \ln  \left( \cos \left( kt \right) 	\right) -2 \ln  \left( \sin \left( kt \right)  \right) }{ \left( \omega+1
		\right) }}
\right].
\end{equation}
\subsection{Without Particle Creation}
We get the normal particle conservation law of standard cosmology in the absence of particle creation, which implies $psi(t)$ = 0.
The following equations are used in continuity Eq. (\ref{9}) for this conservation law: 
\begin{equation}
\label{30}
{\dot{\rho}} + \left [\rho (1+ d) + p (1+ d)\right] \frac{\dot a}{a} = 0~,
\end{equation}
i.e.
\begin{equation}
\label{31}
\frac{{\dot{\rho}}}{\rho} = - (d+1) (1+ \omega)  \frac{\dot a}{a}  ~.
\end{equation}
\begin{equation}
\label{32}
{\dot{\Lambda}}(t) + 8\pi \rho {\dot G(t)} = 0~.
\end{equation}
After the integration of Eq. (\ref{31}), we obtain
\begin{equation}
 \label{33}
 {\rho} = {\rho_{0}} a^ {-(1+ \omega) (d+1)}  ~.
 \end{equation}
 Here $\rho_0$ is a positive constant. From Eq. (\ref{25}) and (\ref{33}), we find,
\begin{equation}
 \label{34}
 {\rho} = \rho_0  \left( a_1  \tan \left( \frac{kt}{2} \right)   
 \right) ^{- \frac{\left( 1+\omega \right)  \left( d+1 \right)}{m} }
   .
 \end{equation} 
After solving Eqs. (\ref{5}), (\ref{32}) and (\ref{34}), we get the gravitational constant $G(t)$ and cosmological constant $\Lambda (t)$:

 \begin{equation}
 \label{35}
 G =   
 -{\frac {d{k}^{2}\cos \left( kt \right) \left( \tan \left( \frac{kt}{2}\right) \right) }{4 m \pi \left( \sin \left( kt
 		\right)  \right) ^{3} \left( 1+\omega \right) \rho_0  \left( a_0
 		 \tan \left( \frac{kt}{2}  \right) \right) ^\frac{{-
 			\left( 1+\omega \right)  \left( d+1 \right) }}{m} \left( 1+ \left( \tan \left( \frac{kt}{2} \right)  \right) ^{2}
 	\right)
 	} }  ,
 \end{equation} 
 
  \begin{equation}
  \label{36}
  {\Lambda} = 
 {\frac {d (d+1){k}^{2}}{2 \left( m \sin \left( kt \right) 
  		\right) ^{2}}}- {\frac {2 d {k}^{2}\cos \left( kt \right) 
  		\tan \left( \frac{kt}{2} \right) }{m \left( 1+\omega \right) \left( \sin \left( kt \right)  \right) ^{3} 
  		\left( 1+ \left( \tan \left( \frac{kt}{2} \right)  \right) ^{2}
  		\right) }}.
    \end{equation}


   \section{Results and Discussions}
   Figure $1(a)$ and $1(b)$ corresponding to the Eq. (\ref{26}) $\&$ Eq. ({34}), portrays the behavior of periodic variation of 
   the energy density ($\rho$) with and without particle creation verses cosmic time (t) for the dimension $5$ and three distinct 
   values of $m$ and $k$. It is observed from the figure $1(a)$ that energy density ($\rho$) with particle creation has Big Rip 
   singularities at the cosmic time $t = \frac{n \pi}{k}$, where $n$ is an integers $(n = 0,1,2,3,4......)$ \cite{ref58}. Since the 
   PVDP is depending on the choice of the values of $k$ and $m$ and we consider only three distinct values so there exists 
   the cosmic singularities corresponding to the different time period as $t = 0, 12.56,25.12 ...... $ for $k = 0.25$, $t = 0, 6.28, 12.56 ...... $ 
   for $k = 0.5$ and $t = 0, 3.14, 6.28 ...... $ for $k = 1$.
   The interesting aspect is that it begins with a large value at stating time in a given cosmic period and decreases to a minimum 
   $\rho$, and then increases again with the growth of time. The minimum energy density occurs at the time $t = \frac{(n+1) \pi}{2 k}$.\\
   
   From the figure $1(b)$, we observe that energy density without particle creation has also periodic singularities at the cosmic time 
   $t = \frac{(2n-1) \pi}{k}$, where $n$ is an integers $(n = 1,2,3,4......)$. In this figure various cosmic singularities are exists for 
   different values of $k$, i.e. $t = 12.56, 37.68, 87.92 ...... $ for $k = 0.25$, $t = 6.28, 18.84, 43.96 ...... $ for $k = 0.5$ and 
   $t = 0, 3.14, 9.42, 21.98 ...... $ for $k = 1$. In beginning of the evolution of the universe, it was infinitely large, indicating 
   the Big-bang scenario. Then it dropped first rapidly, and slowly. It approaches to a smallest value of rho and as time progresses, it 
   increases again. Since $t$ tends to $t_s$, energy density diverge strongly for all values of $k$ and $m$ i.e. $\rho$ tends to infinite 
   so it has Type-I singularity. 
   
   \begin{figure}[H]
   	(a)\includegraphics[width=8cm,height=6cm,angle=0]{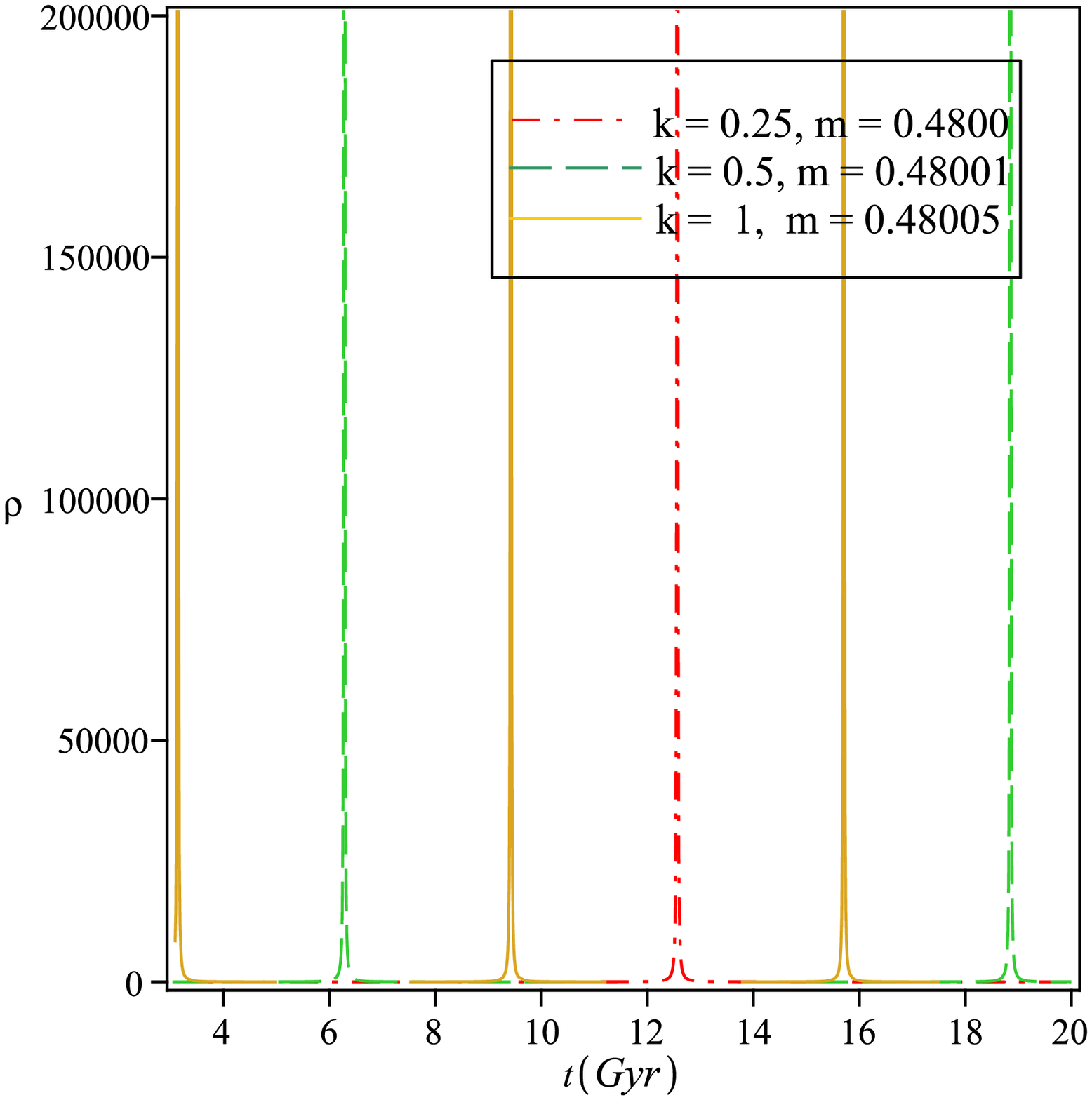}
   	~~~~~~~~~~~~~~(b)\includegraphics[width=8cm,height=6cm,angle=0]{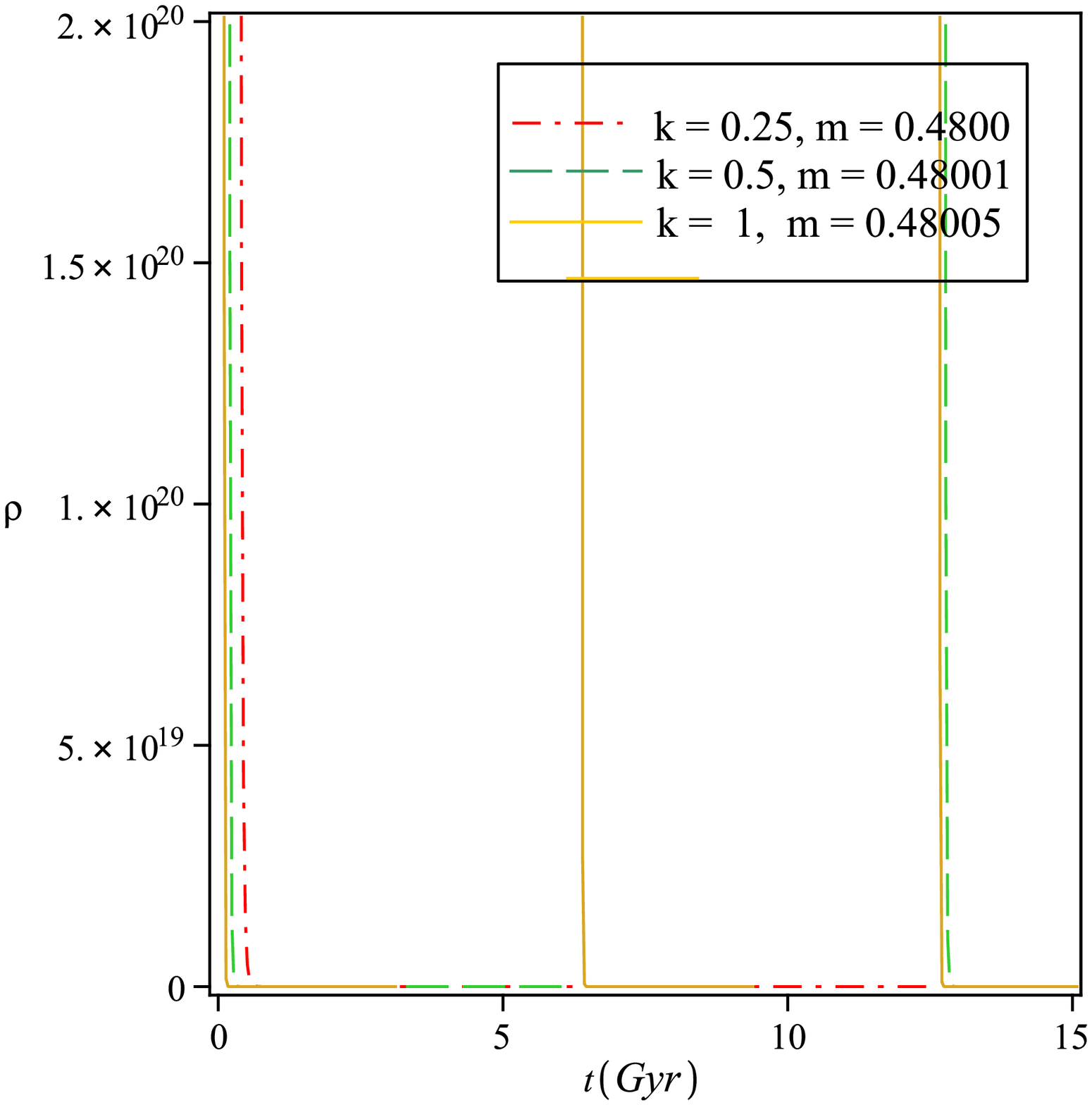}
   	~~\caption{(a) Energy density for particle creation versus cosmic time $t$, (b) Energy density for no particle creation 
   	versus cosmic time $t$ }
   \end{figure}
  
 In this model we analyze the cosmological term $\Lambda$ which will determine the universe's nature. This is plotted in Fig. $2(a)$ and $2(b)$ 
 corresponding to the Eq. (\ref{27}) $\&$ Eq. (\ref{36}) respectively for with and without particle creation. \\
 
 Here, in Figure $2(a)$, the cosmological term $\Lambda$ exhibits a periodic variation with singularity at cosmic time $t = \frac{n \pi}{k}$ 
 (n = 0, 1, 2, 3.....) for all three values of $m$ and $k$ corresponding to FLRW metric. And in $2(b)$ figure cosmological term $\Lambda$ has 
 singularity at cosmic time $t = \frac{(2n-1) \pi}{k}$. We see that within a given cycle cosmological constant $\Lambda$ starts from large 
 positive values and approach to the smallest positive value of $\Lambda$ and again increases with the growth of time i.e. this is a 
 big rip singularity. Thus, the nature of $\Lambda$ in present models is supported by observational evidences.

   
 \begin{figure}[H]
   	\centering
   	(a)\includegraphics[width=7cm,height=6cm,angle=0]{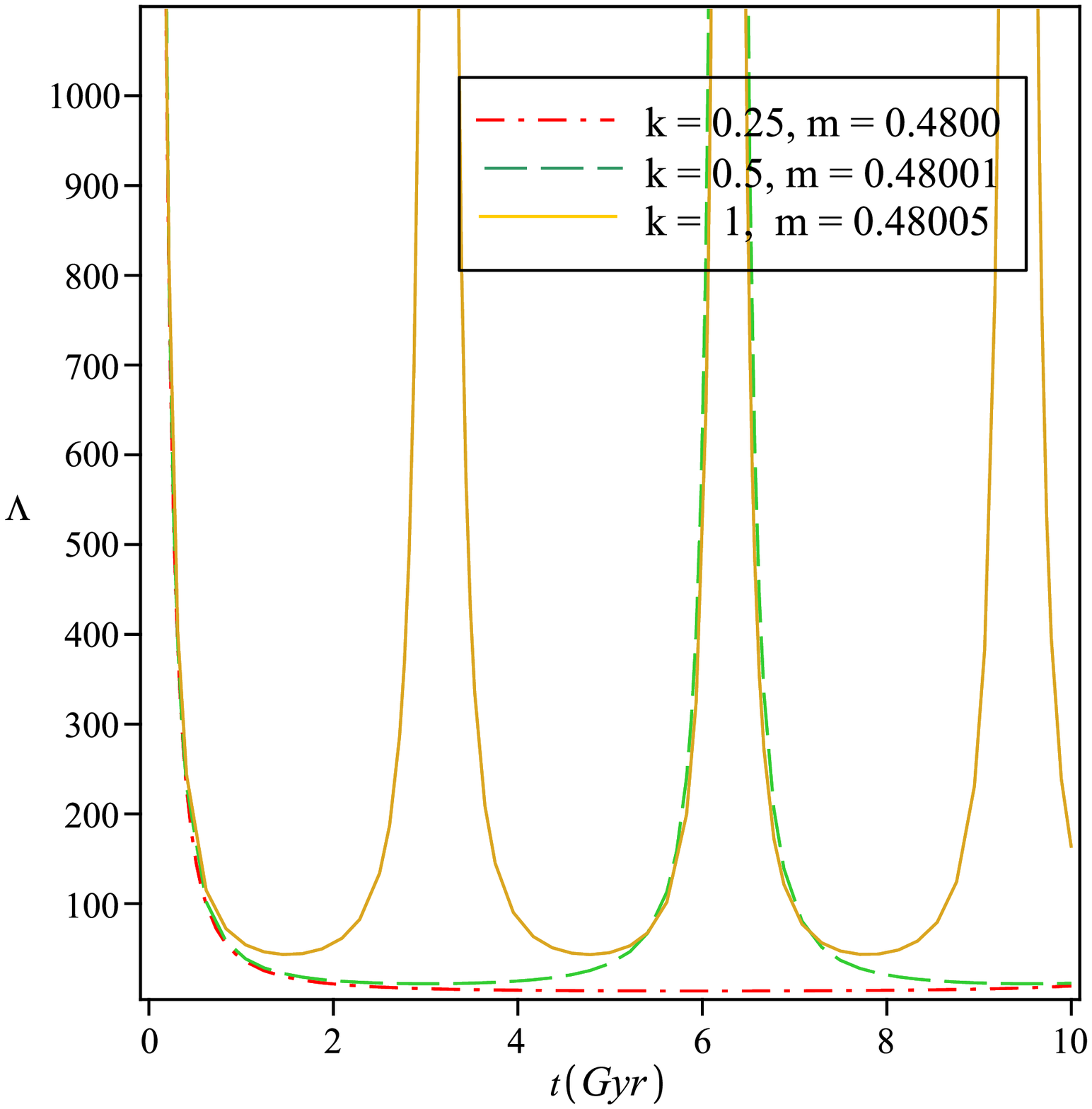}
   	~~~~~~~~~~~~~~(b)\includegraphics[width=8cm,height=6cm,angle=0]{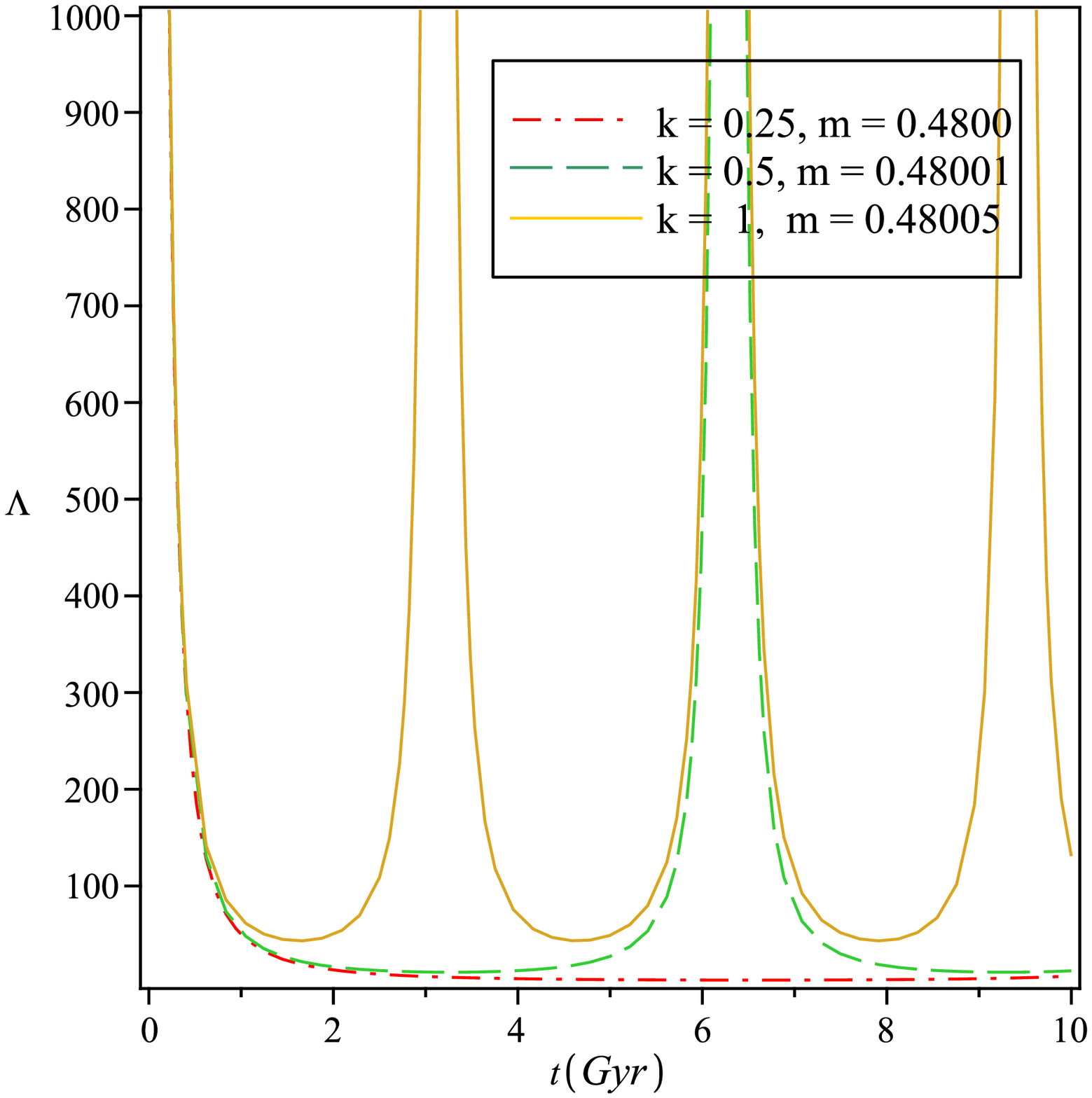}
   	
   	~~\caption{(a) Cosmological constant with particle creation vs. cosmic time $t$, (b) Cosmological constant without particle 
   	creation vs. cosmic time $t$ }
   \end{figure}
 
  Figures $3(a)$ and $3(b)$ demonstrate evolutionary behavior of pressure $p$ with respect to time $t$ for all values of $k$ and $m$ 
  corresponding of FLRW metric for dimension $5$ (d=5). From the figure $3(a)$ it is ascertained that the pressure have a periodic 
  variation with singularity at cosmic time $t = \frac{n \pi}{k}$ (n = 0, 1, 2, 3.....). Here, we find a repeated cyclic pattern, where 
  at the beginning pressure decreases from large positive value to large negative values and it keeps flowing on. \\
 
 And in figure $3(b)$ nature of pressure is also periodic and contain singularity at cosmic time  $t = \frac{(2n-1) \pi}{k}$, where $n$ is 
 an integers $(n = 1,2,3,4......)$ i.e. $t = \frac{\pi}{k}, \frac{3\pi}{k},\frac{5\pi}{k}.......$. It is negative throughout 
  the evolution. From the figure we observe that pressure is decreasing function of  time $t$ $\&$ it begins from an large negative value and 
  reaches zero. We find this is repeated cycle pattern. We see that pressure $p$ is high negative at an early stage however it decreases as 
  time will increase.

   \begin{figure}[H]
   	\centering
   	(a)\includegraphics[width=7cm,height=6cm,angle=0]{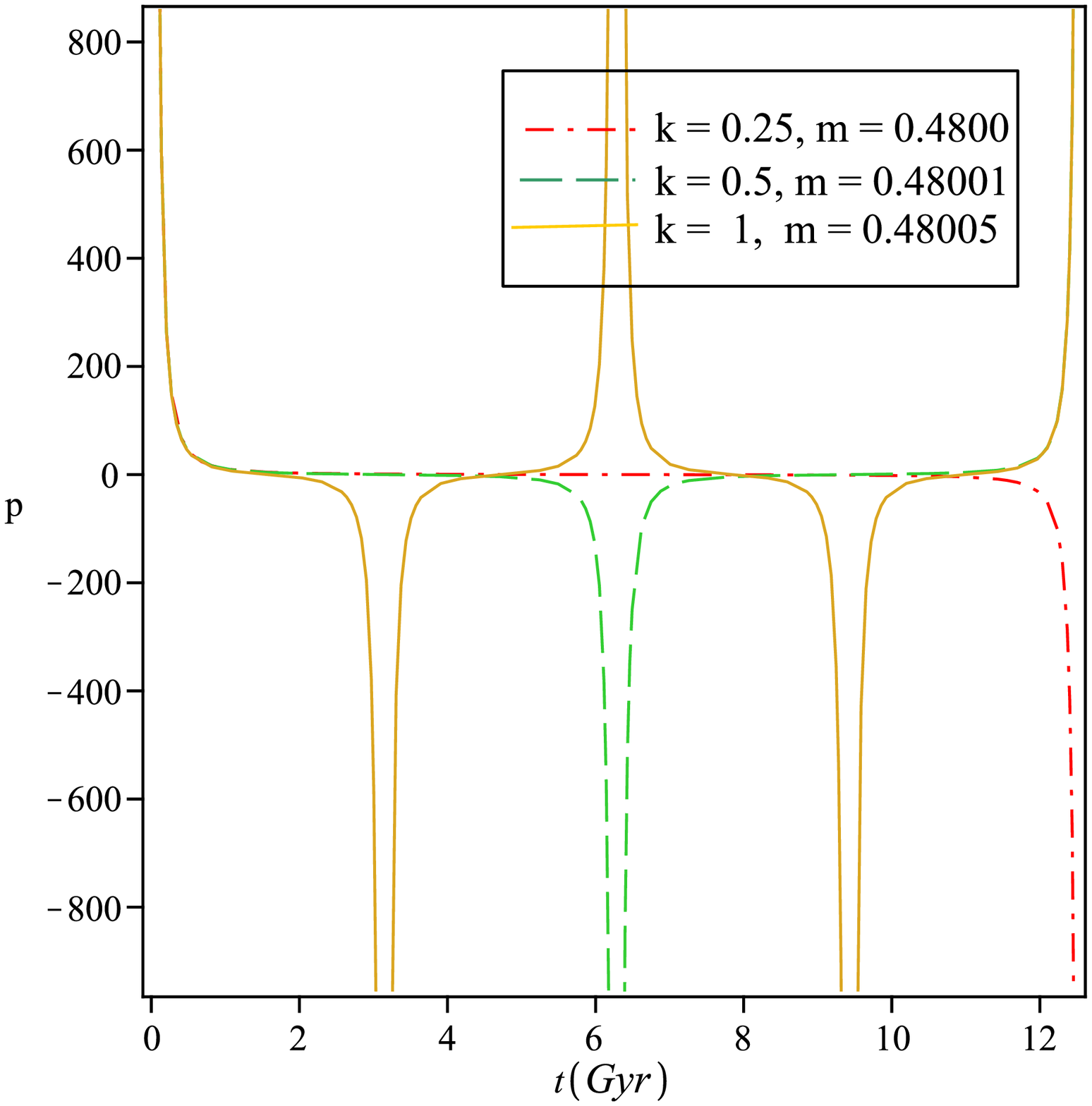}
   	~~~~~~~~~~~~~(b)\includegraphics[width=8cm,height=6cm,angle=0]{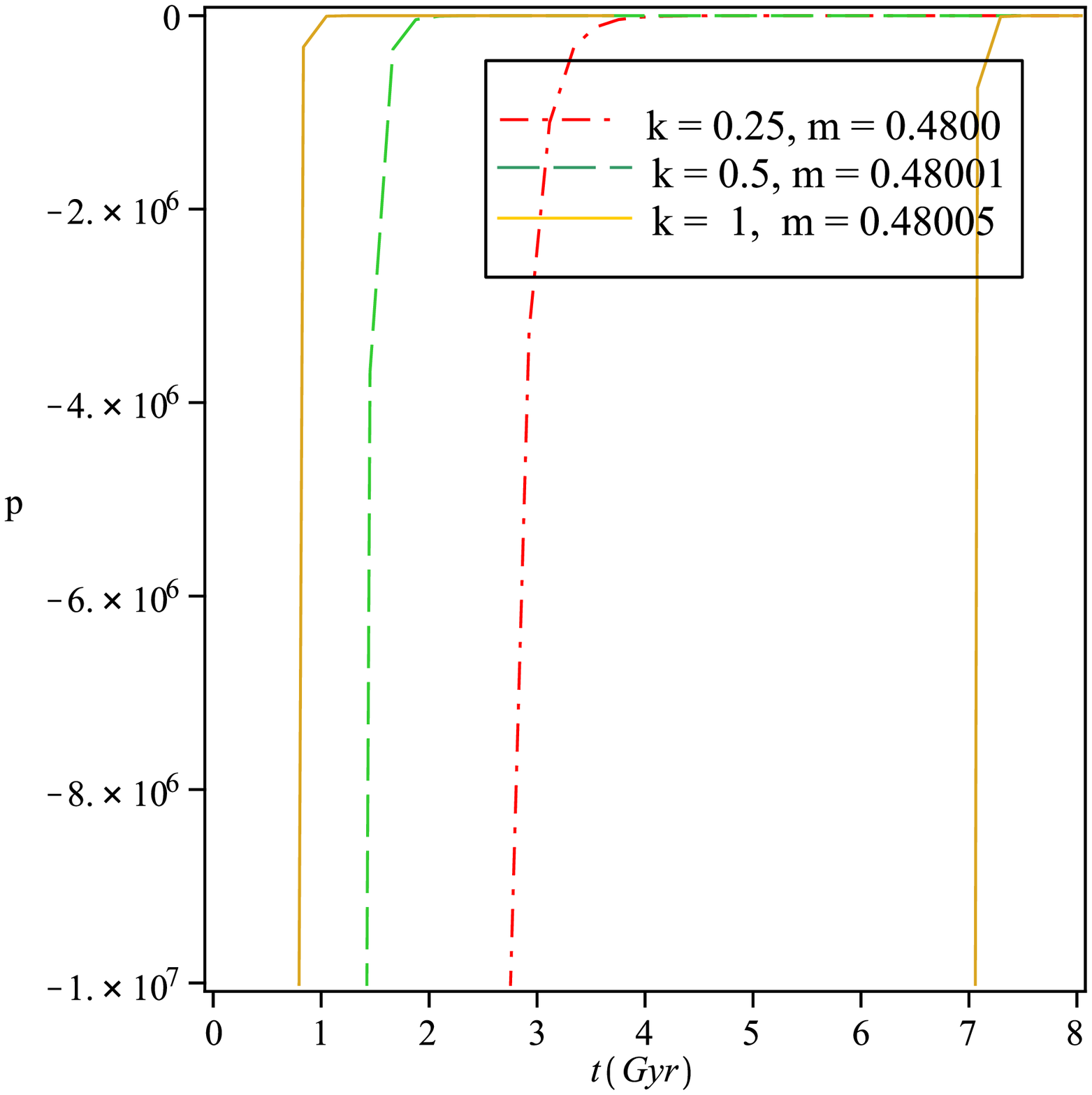}
   	
   	~~\caption{(a) Pressure for particle creation versus cosmic time $t$, (b) Pressure for no particle creation versus cosmic time $t$ }
   \end{figure}
   
    Figure $4(a)$ shows periodic variation behavior of particle creation $\psi$ with cosmic time $t$ for all three values of $m$ and $k$.
   From the figure it is observed that particle creation ($\psi$) has 'Big Rip' singularities at the cosmic time $t = \frac{n \pi}{k}$,  
   i.e. $t = 0, \frac{\pi}{k}, \frac{2\pi}{k},\frac{3\pi}{k}.......$, where $n$ is an integers $(n = 0,1,2,3,4......)$. We can see from the 
   figure that the nature of time dependent particle creation $\psi$ is consistent with the standard observational data.\\
 
 In the case of no particle creation, Figure $4(b)$ demonstrate the periodic variation of gravitational constant with singularities at the 
 cosmic time $t = \frac{(2n-1) \pi}{k}$. From the figure, we can see that the gravitational constant is only an increasing function for all 
 values of $m$ and $k$ as expected.
    
   \begin{figure}[H]
   	\centering
   	(a)\includegraphics[width=7cm,height=6cm,angle=0]{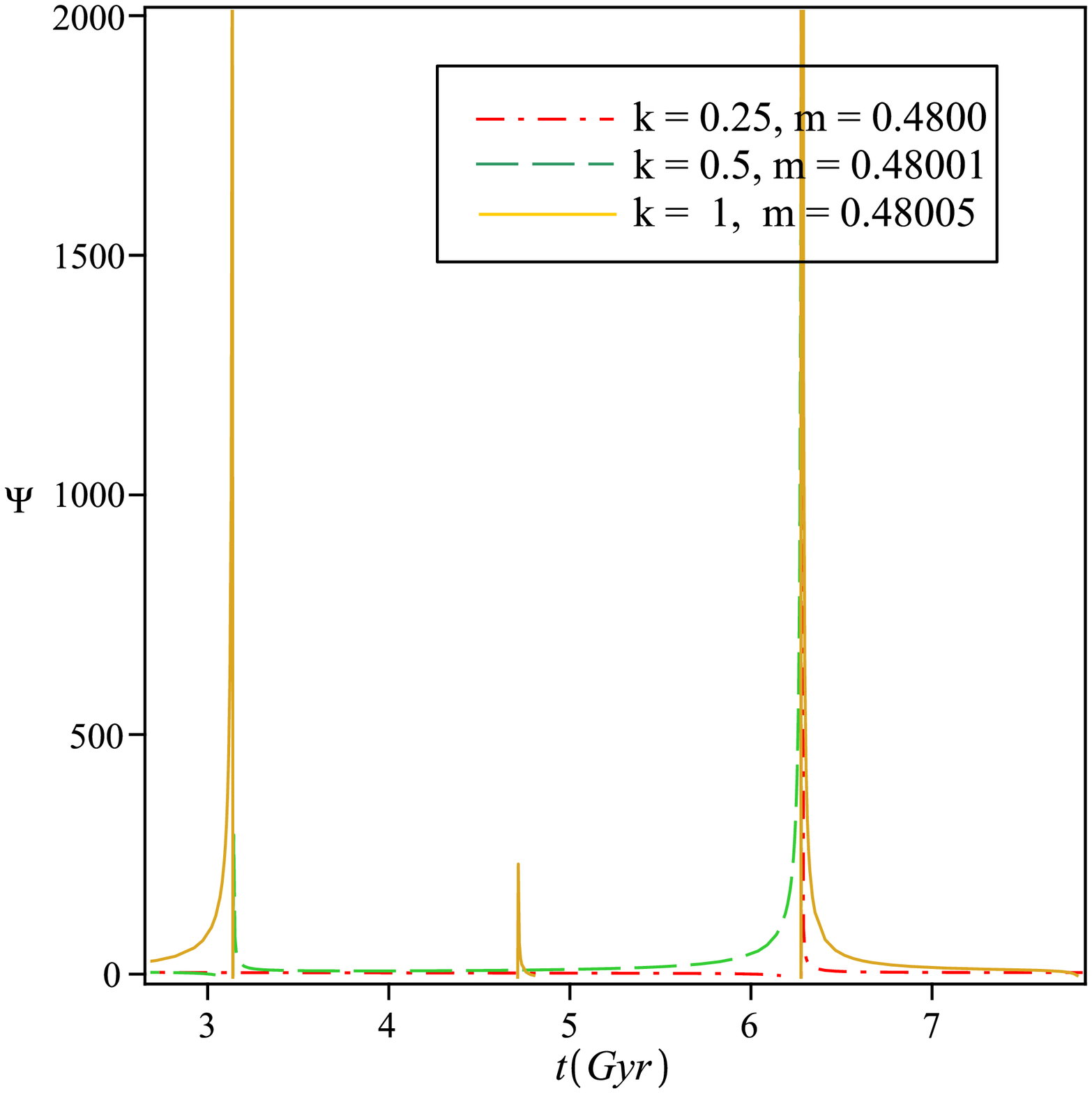} 
   	~~~~~~~~~~~~(b)\includegraphics[width=8cm,height=6cm,angle=0]{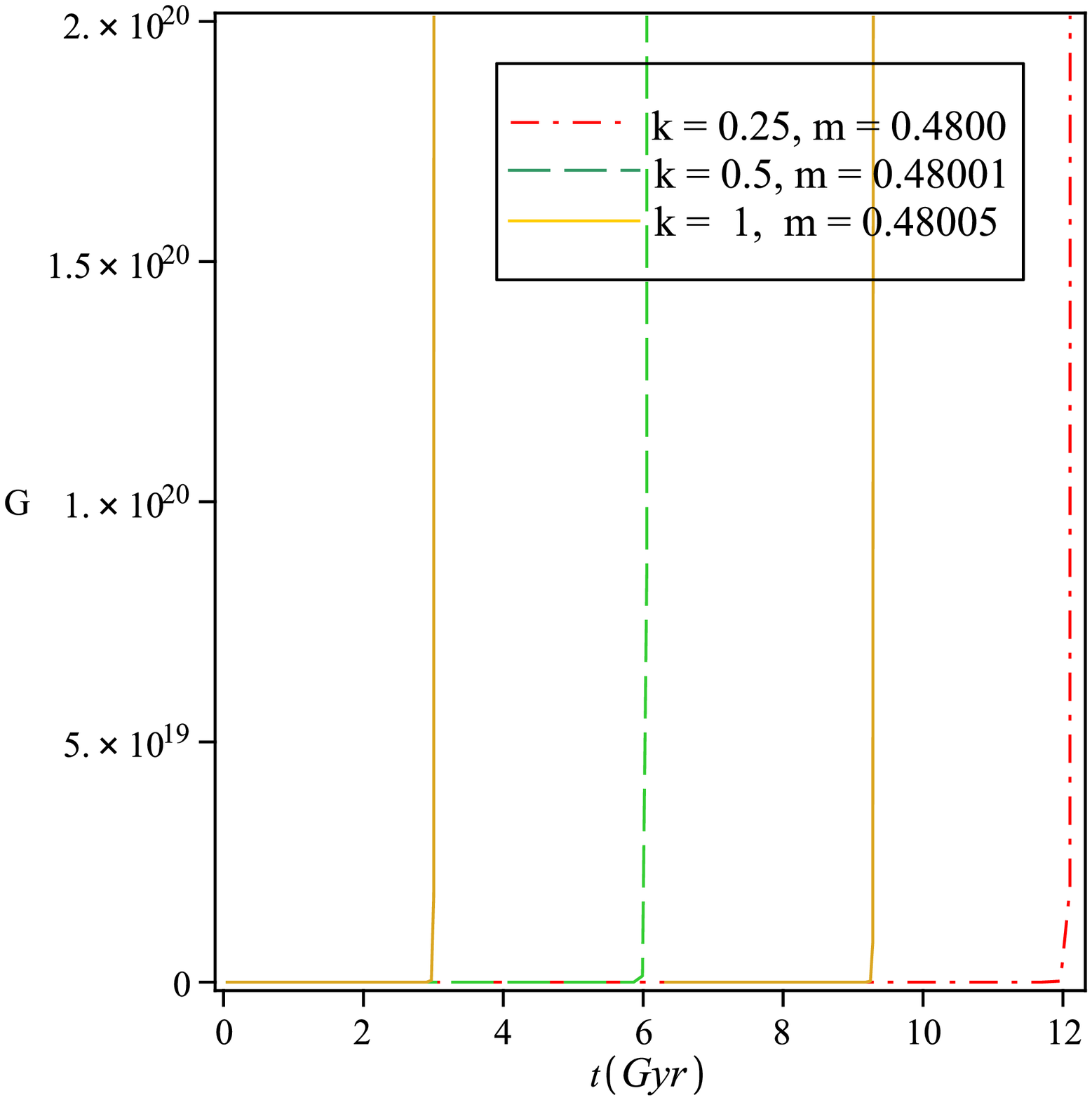}
   	~~\caption{(a) Particle creation versus cosmic time $t$, (b)  Gravitational constant versus cosmic time $t$ }
   \end{figure}
   
    Figure $5$ demonstrates entropy creation with cosmic time $t$ for different values of $k$ and $m$. It has periodic singularity at 
    cosmic time  $t = \frac{n \pi}{k}$ for all values of $k$ and $m$, here $n$ is an integers $(n = 1,2,3,4......)$ i.e. $t = \frac{\pi}{k}, 
    \frac{2\pi}{k},\frac{3\pi}{k}.......$.
    
     \begin{figure}[H]
    	\centering
    	(a)\includegraphics[width=7cm,height=6cm,angle=0]{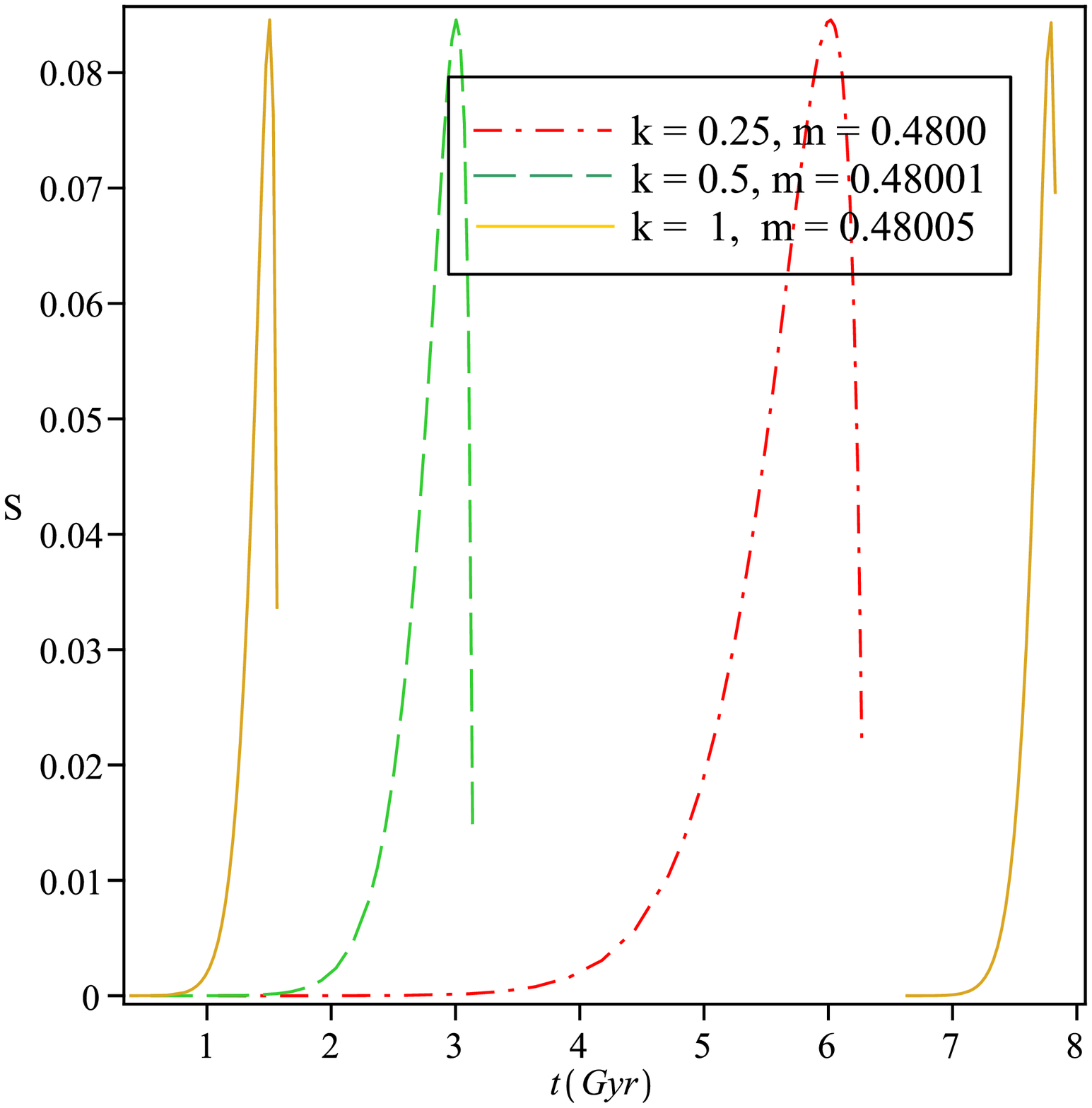}
    	~~\caption{Entropy production versus cosmic time $t$ }
    \end{figure}

 \section{Kinematic Test}
 Now, as suggested in the preceding section, we derive some kinematic relationships of the model. 
 
  \subsection{The Density Parameter}
  
 The density parameters for matter $ \Omega_{matt}(t)$ and vacuum $ \Omega_{\Lambda}(t)$ are
   \begin{equation}
   \label{37}
  \Omega_{matt}(t) = \frac{\rho}{\rho_c},
   \end{equation}
  i.e. 
   \begin{equation}
   \label{38}
   \Omega_{matt}(t) = 
     =  \frac{2 m ~ cos(kt)}{(\omega +1) (d+1)}.
   \end{equation}
 
   \begin{equation}
   \label{39}
   \Omega_{\Lambda}(t) = \frac{\rho_\Lambda}{\rho_c}  = 1 -   \frac{2 m ~ cos(kt)}{(\omega +1) (d+1)},
   \end{equation}
 where $\rho_c = \frac{d(d+1)H^{2}}{16 \pi G}$.
 
 From equations (\ref{38}) and (\ref{39}), total density parameter $ \Omega_{T}$ is
 \begin{equation}
 \label{40}
 \Omega_{T} =  \Omega_{matt} + \Omega_{\Lambda} = 1 ~.
 \end{equation}
The inflationary scenario favors this solution which is equivalent to the normal Einstein gravity. The preliminary outcomes, as per 
the high redshift of Supernovae and CNB, indicate that the universe may accelerate, with a dominant contribution to its energy density 
being received in the form of a cosmological constant. Wide range of values of $\Omega_{matt. 0}$ and $\Omega_{\Lambda 0}$ ( the present 
cosmic matter and vacuum energy density parameters) has been offered by the recent measurements. SNe Ia observation together with the total 
 energy density constraints from CMB \cite{ref65} and combined gravitating lens and stellar dynamical analysis \cite{ref66} contribute 
 to $\Omega_{matt. 0} \sim 0.3$ and $\Omega_{\Lambda 0} \sim  0.7$.

  \subsection{Proper Distance Redshift}

$d(z) = a 0 r(z)$ is the acceptable distance between the source and observer, where $r(z)$ is the radial distance of the target at light 
emission in terms of redshift, such as 
  
  \begin{equation}
  \label{41}
  r{(z)} =  \int_{t}^{t_0} \frac{dt}{a(t)},
  \end{equation} 
  
  \begin{equation}
 \label{42}
 r{(z)} =  \int_{0}^{z} \frac{dz}{a_0 H(z)},
   \end{equation} 
   Hence
   \begin{equation}
   \label{43}
   d{(z)} =  \int_{0}^{z} \frac{2 m dz}{k a_0 (1+z) (1+\frac{1}{ (1+z)^{2m}} ) }.
   \end{equation} 
   
   
    \begin{figure}[H]
    	\centering
    	(a)\includegraphics[width=7cm,height=6cm,angle=0]{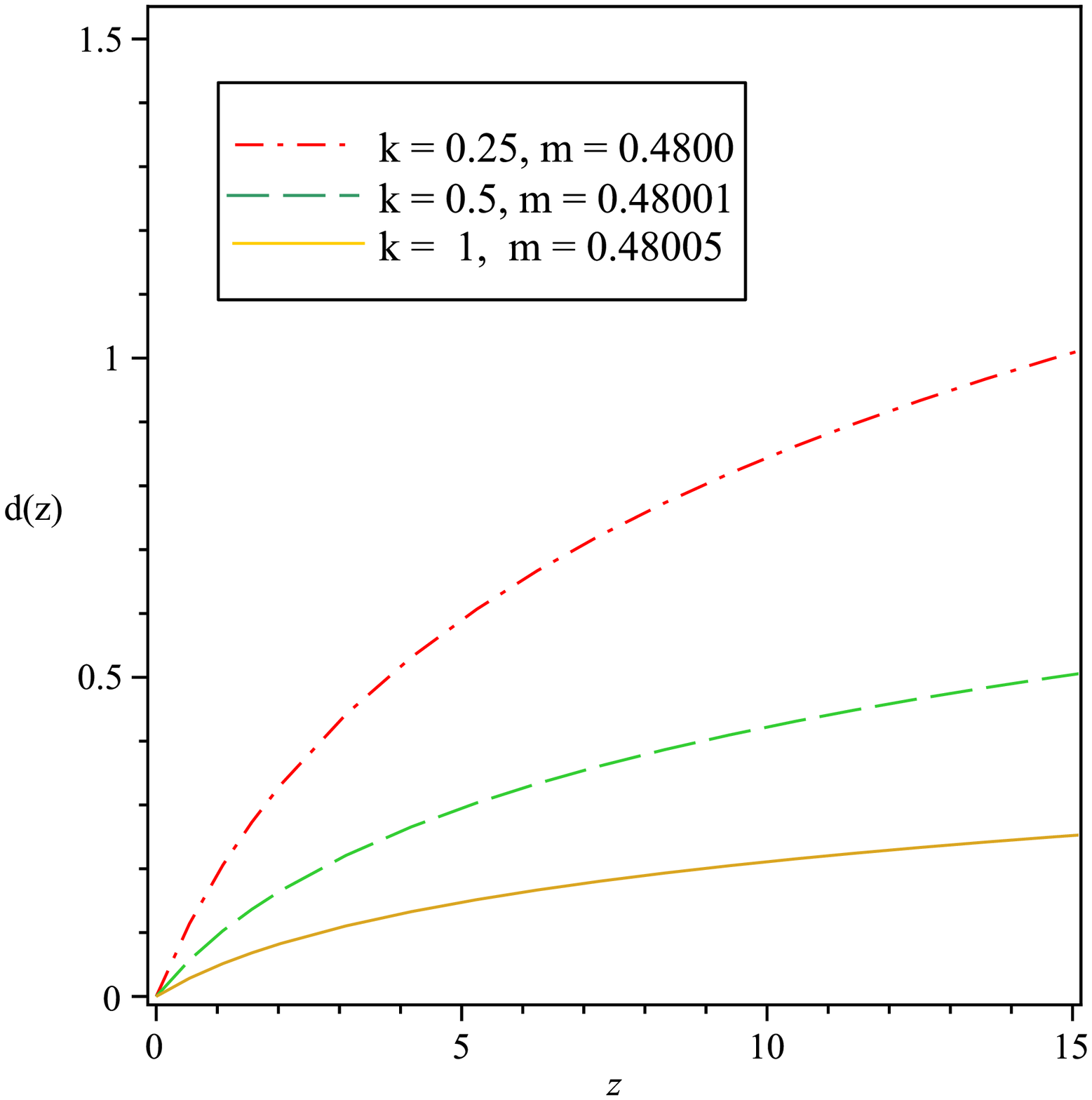}
    	~~\caption{Proper Distance vs Redshift z }
    \end{figure} 
    
  For some selected values of $m$ and $k$ the proper distance as a function of redshift shown in figure 6. We observed that particle creation 
  gives rise to proper distance.
  
  \subsection{Angular distance Redshift}  
   The another important Kinematic test is angular diameter distance redshift which is denoted by $d_A$. This is the ratio of physical 
   transverse size of an object to its angular size (in radians). It is given by in term of $z$ such as 
  \begin{equation}
  \label{44}
  d_{A} =  \frac{d(z)}{(1+z) } .
  \end{equation} 
 Using the Eq. (\ref{43}), angular distance redshift is    
  \begin{equation}
  \label{45}
  d_{A} =  \frac{1}{(1+z) } \int_{0}^{z} \frac{2 m dz}{k a_0 (1+z) (1+\frac{1}{ (1+z)^{2m}} ) }  .
  \end{equation}  
In Figure $7$, we draw the variation of angular distance vs redshift $z$ for three values of $m$ and $k$. This shows that particle creation enhance the 
angular distance. The angular diameter distance initially increases with increasing $z$ and gradually starts to decrease for all values of 
$k$ and $m$ ($k = 0.25$, $m= 0.4800$ and $k = 0.5$, $m= 0.48001$ $k = 1$, $m= 0.48005$).

   \begin{figure}[H]
   	\centering
   	(a)\includegraphics[width=7cm,height=6cm,angle=0]{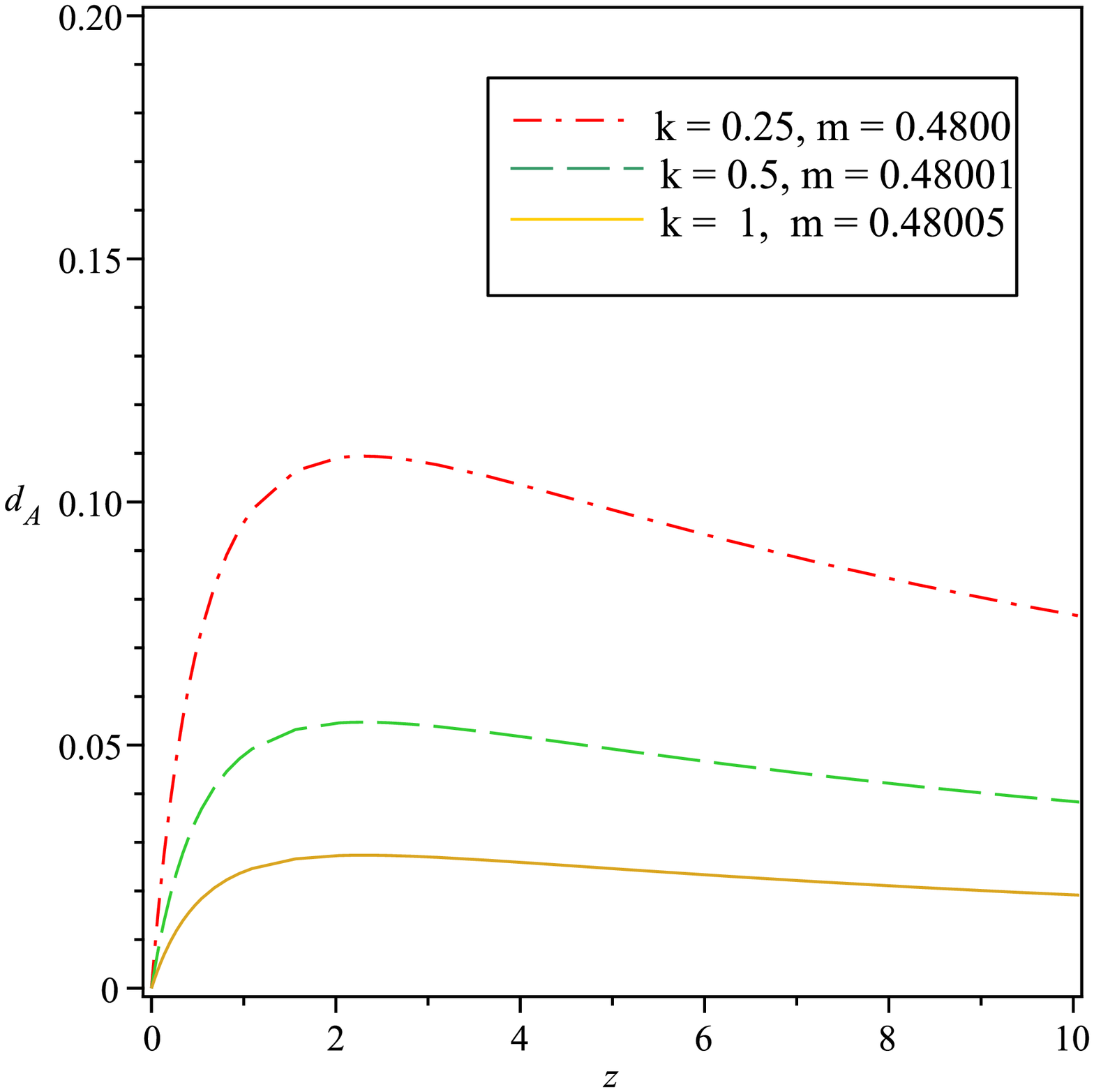}
   	~~\caption{Angular distance vs Redshift z }
   \end{figure} 
   
   
 \subsection{Luminosity Distance Redshift}
This is pointed out by the observations of Ia Supernova \cite{ref5,ref6} that universe expansion is in the accelerating phase. In this 
 Redshift plays an important role. luminosity distance versus time $t$ \cite{ref67,ref68 } is an important observational tool to study the 
 evolution of the universe. The expansion of the universe causes the light which is emitted by the stellar object to get redshifted. 
 The concept of distance states the explanation of the expanding universe which is linked with the observations. It has been defined in 
 various ways in the literature. Specifically, luminosity distance is the distance defined by the luminosity of a stellar object and has 
 an important role in astronomy. One can also compute the rate of expansion of the universe through the observational measurements of 
 luminosity distance. Here luminosity distance $D_L$ in terms of redshift has been derived.\\
 
 Expression for the Luminosity distance determining flux of the source with red shift $z$ is 
 \begin{equation}
 \label{46}
 D_{L} = a_0 c (1+z) \int_{t}^{t_0} \frac{dt}{a(t)} ,
 \end{equation}
 where $c$ expresses speed of light and $a_0$ shows the present value of the scale factor.
 \begin{equation}
 \label{47}
 = c (1+z) \int_{0}^{z} \frac{dz}{H(z)} .
 \end{equation}
 Since $H(z) = \frac{k (1+z)}{2m} \left(1+ \frac{1}{ (1+z)^{2m}} 
 \right)$, so Eq. (\ref{49}) become, 
 \begin{equation}
 \label{48}
 D_{L} =  c (1+z) \int_{0}^{z}  \frac{2 m  dz}{k (1+z) (1+ \frac{1}{ (1+z)^{2m}} ) } .
 \end{equation}
 
  \begin{figure}[H]
  	\centering
  	(a)\includegraphics[width=7cm,height=6cm,angle=0]{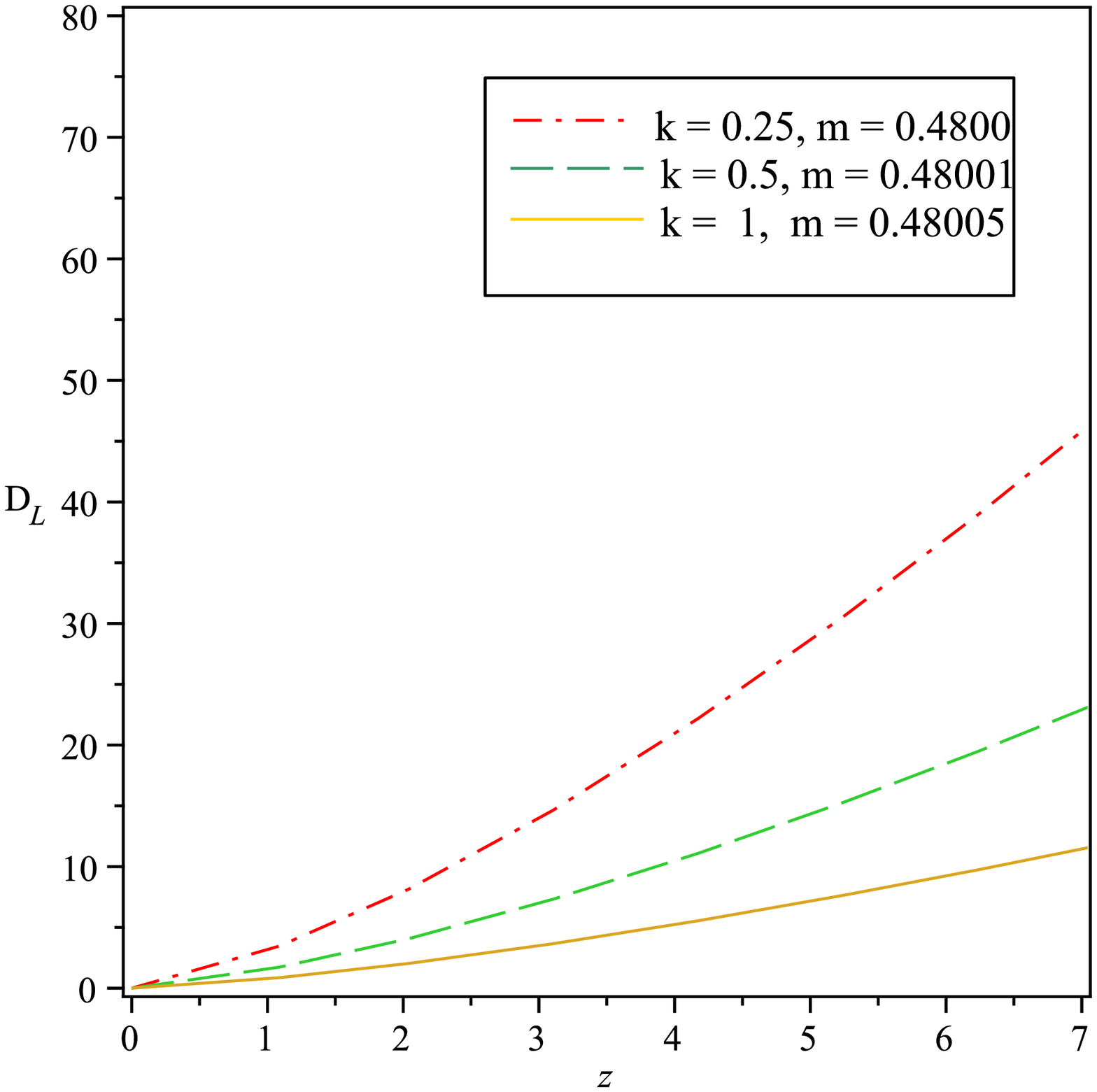}
  	~~\caption{Luminosity distance vs Redshift z }
  \end{figure}  
  
  
\subsection{Apparent Magnitude}
  The distance modulus $\mu$ which is difference of apparent magnitude and absolute magnitude and related to the luminosity distance define as
  \begin{equation}
  \label{49}
  \mu = m - M = 25 + 5 log_{10} \left(\frac{D_{L}}{ Mpc}\right) ,
  \end{equation}
  where $m$ denotes apparent magnitude and $M$ denotes absolute magnitude respectively.\\
  For finding the small red shift, we use the following equation of $D_L$ given as 
  \begin{equation}
  \label{50}
  D_{L} =  \left(\frac{cz }{H_0}\right) ,
  \end{equation}
  There are a lot of supernova of low red shift whose apparent magnitudes are known. we have find absolute magnitude $M$ of 
  Type Ia supernova(SNIa) \cite{ref69,ref70} by using $z = 0.026$ and $m = 16.08$ in Eq. (\ref{51}) as follows: 
  \begin{equation}
  \label{51}
  M =  5 log_{10} \left(\frac{H_0}{0.026 c}\right) - 8.92 ,
  \end{equation}
  From this we obtain apparent magnitudes $m$ as
  
  \begin{equation}
  \label{52}
  m = 16.08 + 5 log_{10} \left(\frac{D_{L} H_0}{0.026 c}\right) ,
  \end{equation}
  \begin{equation}
  \label{53}
  m = 16.08 + 5 log_{10} \left(\frac{H_0 (1+z)}{0.026} \int_{0}^{z} \frac{dz}{H(z)} \right) ,
  \end{equation}
  \begin{equation}
  \label{54}
  m = 16.08 + 5 log_{10} \left(\frac{H_0 (1+z)}{0.026} \int_{0}^{z} \frac{2 m dz}{k a_0 (1+z) (1+ \frac{1}{(1+z)^{2m}} ) } \right) ,
  \end{equation}
  
 \begin{figure}[H]
 	\centering
 	(a)\includegraphics[width=7cm,height=6cm,angle=0]{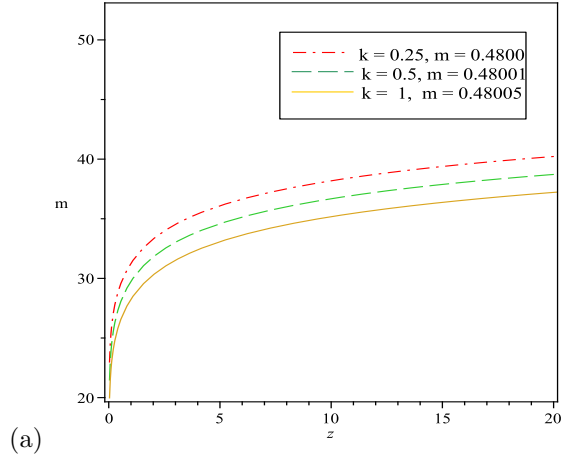}
 	~~\caption{Apparent magnitude vs Redshift z }
 \end{figure} 
 Figure $9$ depicts apparent magnitude versus redshift $z$ for the observational values of (m, k).
 
 \subsection{Age of the universe versus redshift $z$} 
 The age of the universe is define as
 \begin{equation}
 \label{55}
 t_{0} = \int_{0}^{t_{0}} dt = \int_{0}^{\infty}\frac{dz}{H(z) (1+z)},
 \end{equation}
 Since $H(z) = \frac{k (1+z)}{2m} \left(1+ \frac{1}{ (1+z)^{2m}} 
 \right)$, so Eq. (\ref{55}) become
 \begin{equation}
 \label{56}
 t_{0} = \int_{0}^{\infty} \frac{2 m dz}{k (1+z)^2 (1+\frac{1}{ (1+z)^{2m}} ) },
 \end{equation}
 
 \begin{figure}[H]
 	\centering
 	(a)\includegraphics[width=7cm,height=6cm,angle=0]{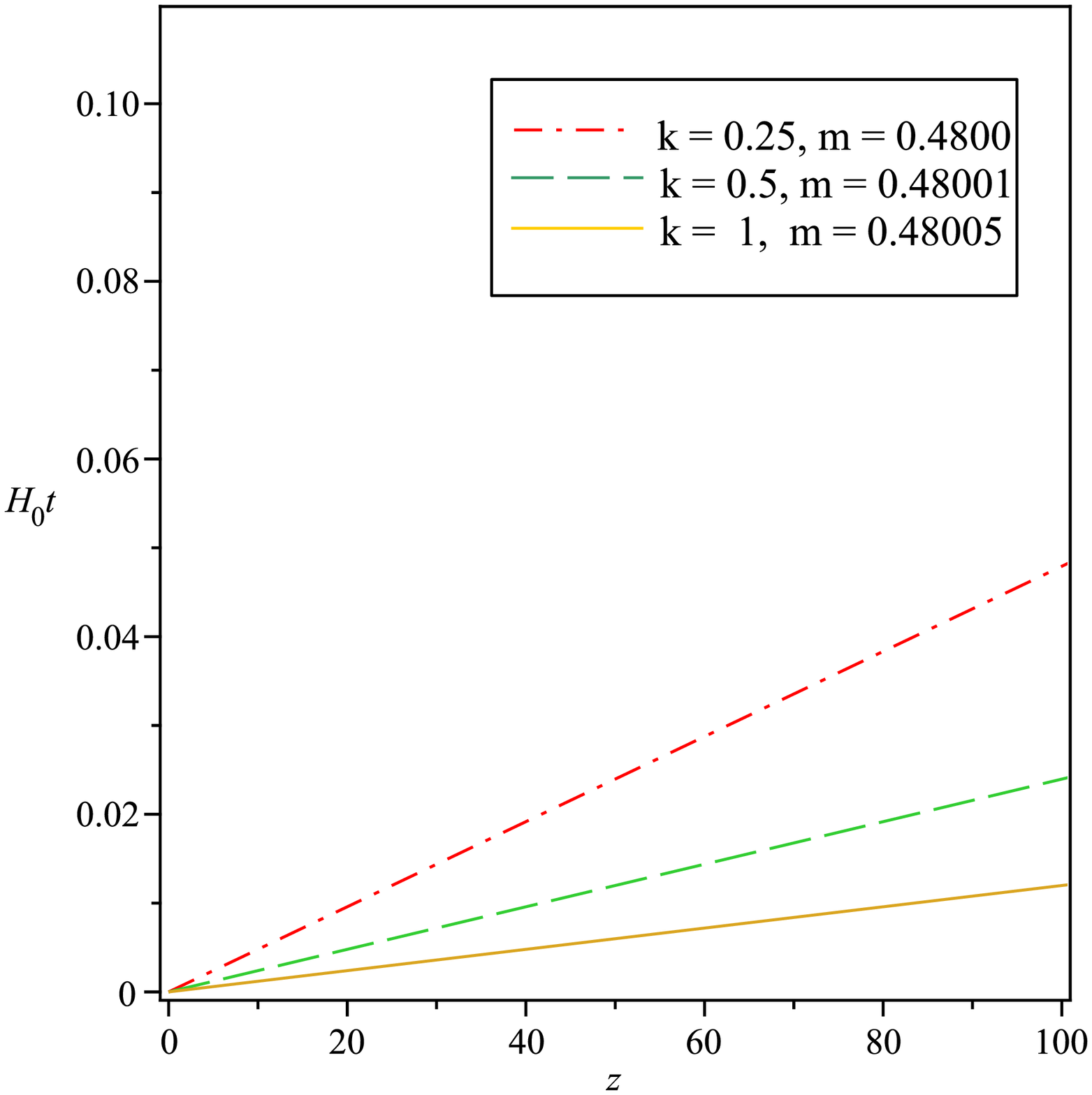}
 	~~\caption{Age of the universe versus redshift $z$ }
 \end{figure}
 
In Fig. $10$, the curves are drawn for $H_0$ t with respect to redshift $z$. It shows that $H_0$ t is in increasing order with 
 increasing $z$ for all values of $k$ and $m$ ($k = 0.25$, $m= 0.4800$ $\&$ $k = 0.5$, $m= 0.48001$ and $k = 1$, $m= 0.48005$).
 Consequently, this gives the universe age $t_{0}$ as 10.34 Gyrs and 12.36 Gyrs. The age of our universe, according to WMAP info, 
 is approximately 13.73 Gyrs. So, the closest available theoretical value of $t_{0}$ is Gyrs $12.36 .$ However the drawn curves are 
 best suited and in full agreement with observed results for $H_{0}=69.2$. 
 
  
 \subsection{Look-back Time Redshift} 
  
 The look-back time is the difference between the age of universe at the present time $z=0$ and the age of the universe when a specific 
 right ray was produced at redshift $z$ \cite{ref71}. For a given redshift $z,$ scale factor $a$ is related to the $a_{0}$ by the relation 
 $a=a_{0}(1+z)^{-1}$. since $1 \leq(1+z)<\infty$ is the inverse of the scale factor $0<a \leq 1,$ for both $z$ and $(1+z)$ are potentially 
 misleading and nonlinear functions of fundamental quantities.\\

 In terms of the scale factor $a$ and $H = d~ln(a)/dt$, the look-back time is
  
  \begin{equation}
  \label{57}
  t_{L} = \int_{t}^{t_0} dt ~~= \int_{1}^{a} \left(\frac{dt}{d ln(a)}\right) d ln(a)
  \end{equation}

  \begin{equation}
  \label{58}
  t_{L} = \int_{1}^{a} \frac{da}{a H(a)} 
  \end{equation}
  
  The look back time  can be written in terms of redshift (z):
  \begin{equation}
  \label{59}
  t_{L} = \int_{t}^{t_0} dt ~~~~ = \int_{z}^{0} \left(\frac{dt}{dz}\right) dz 
  \end{equation}
  
  Using the  Equation $a = a_0 (1+z)^{-1}$ to determine $dt / dz$ as: 
  
  \begin{equation}
  \label{60}
  \frac{dz}{dt} = \frac{-1}{a^{2}}\frac{da}{dt} ~~=  \frac{1}{a}\left(\frac{\dot a}{a}\right) = - (1+z) H
  \end{equation}
  
  \begin{equation}
  \label{61}
  t_{L} = \int_{0}^{z} \frac{dz}{(1+z) H(z)}  
  \end{equation}
  \begin{equation}
  \label{62}
  t_{L} = \int_{0}^{z} \frac{2 m dz}{k a_0 (1+z)^2 \left(1+\frac{1}{ (1+z)^{2m}} \right) }  
  \end{equation}
  \begin{figure}[H]
  	\centering
  	\includegraphics[width=7cm,height=6cm,angle=0]{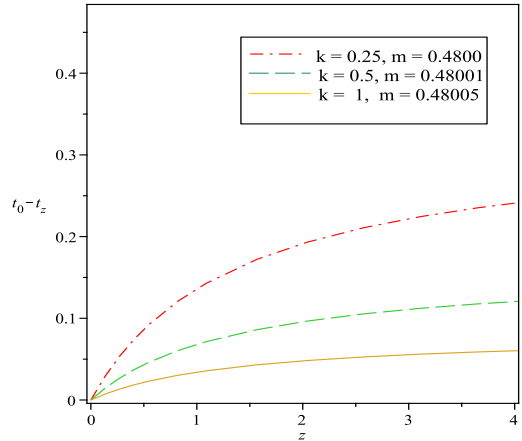}
  	~~\caption{Look-back time versus redshift $z$ .}
  \end{figure}
 We have observed in Fig $11$ that the look-back time is increases with redshift $(z)$ for all values of $k$ and $m$.


\section{Concluding remarks}
In the present work, we consider a periodically varying deceleration parameter to reconstruct the cosmic history. We also analyzed the 
mechanism of particle creation admitting variations of the cosmological constant $\Lambda$ and the gravitational 
constant $G$ for higher-dimensional FLRW space-times.\\

Apparently, the universe's dynamic properties have been periodically produced by the PVDP. Within the specified cosmic duration and 
defined by the cosmic frequency parameter of the model, the energy density $\rho$ $\&$ pressure $p$ vary cyclically. The magnitude of these physical 
parameters turns infinitely large at some finite time. This behavior contributes to the singularity of type I as categorized 
by Nojiri et al  \cite{ref72}. There seems to be a Big Rip singularity at a certain finite period during the cosmic replication of the process since 
$a \rightarrow \infty, \rho \rightarrow \infty$ and $|p| \rightarrow \infty $.\\

We notice that for the particle creation, behavior of the parameters repeat after the time period $t=\frac{n \pi}{k}$ i.e. the big Rip occurs 
periodically after a time gap. Similar behavior of the parameters is also for without particle creation for the time period $t=\frac{(2n-1) \pi}{k}$ . \\

Here we observed that particle creation $\psi$ and entropy $S$ has periodic variation with singularity at the cosmic time $t=\frac{n \pi}{k}$ i.e. 
 these have Big-Rip singularity. We also note that 
 gravitational constant $G$ has cyclic nature with singularity at the cosmic time $t=\frac{(2n-1) \pi}{k}$. From the figures we analyze that 
 cosmological constant $\Lambda$ has singularity for both cases particle creation and without particle creation.  Then, we investigated 
 the periodic variation of several cosmological quantities such as energy density$\rho$, 
 pressure $p$, particle creation rate $\psi$, entropy $S$, etc. which have Big Rip singularities.\\

We have also explored some more parameters through some kinematics tests like Proper distance, Angular distance, 
Luminosity distance $(D_L)$, Apparent Magnitude $(m)$, Age of the Universe, and Look back time $(t_L)$ with respect
to redshift $(z)$. These results are found to be compatible with the current observations.\\

Hence, our constructed model and their solutions have good agreement with the observational data and physically acceptable. Therefore, for more 
understanding of the characteristics of the particle creation in our universe’s evolution within the framework of FLRW metric, 
the solution demonstrated in this paper may be helpful.





\begin{thebibliography}{99}
\bibitem {ref1}
P.M. Garnavich, et al., Constraints on cosmological models from Hubble Space Telescope observations of High-z Supernovae, 
Astrphys. J. {\bf 493}, L53 (1998).
\bibitem {ref2}
P.M. Garnavich, et al., Supernova limits on the cosmic equation of state, Astrphys. J. {\bf 509}, 74 (1998).
\bibitem {ref3}
S. Perlmutter, et al., Measurements of the cosmological parameters $\Omega$ and $\Lambda$ from the first seven Supernovae at $z \geq 0.35$, 
Astrophys. J. {\bf 483}, 565 (1997).
\bibitem {ref4}
S. Perlmutter, et al., Discovery of a Supernova Explosion at Half the Age of the Universe and its Cosmological Implications, 
Nature {\bf 391}, 51 (1998). 
\bibitem {ref5} 
S. Perlmutter, et al., Measurements of $\Omega$ and $\Lambda$ from 42 high-redshift supernovae, Astrophys. J. {\bf 517}, 565 (1999).
\bibitem{ref6}
A.G. Riess, et al., Observational Evidence from Supernovae for an Accelerating Universe and a Cosmological Constant, Astron. J. 
{\bf 116}, 1009 (1998).
\bibitem {ref7}
B.P. Schmidt, et al., The High-Z Supernova Search: Measuring Cosmic Deceleration and Global Curvature of the Universe Using 
Type Ia Supernovae, Astrophys. J. {\bf 507}, 46 (1998).	
\bibitem {ref8} 
P. De Bernardis, P.A.R. Ade, J.J. Bock, et al., Nature {\bf 404}, 955 (2000).
\bibitem {ref9}
C.L. Bennett, et al., First Year Wilkinson Microwave Anisotropy Probe (WMAP) Observations: Preliminary Maps and Basic Results, 
Astrophys. J. Suppl. {\bf 148}, 1 (2003).
\bibitem {ref10}  
D.N. Spergel, et al., First Year Wilkinson Microwave Anisotropy Probe (WMAP) Observations: Determination of Cosmological Parameters, 
Astrophys. J. Suppl. {\bf 148}, 175 (2003). 
\bibitem {ref11}
M. Tegmark, et al., The three dimensional power spectrum of galaxies from the sloan digital sky survey, Astrophys. J.
{\bf 606}, 702 (2004).
\bibitem {ref12} 
P. Garg, A. Pradhan, R. Zia, and Mohd. Zeyauddin, Decelerating to accelerating scenario for Bianchi type-II string
Universe in f(R, T ) gravity theory, Int. J. Geom. Methods Mod. Phys. {\bf 17}, 2050108 (2020).
\bibitem {ref13}
E. Komatsu, et al., Five-Year Wilkinson Microwave Anisotropy Probe (WMAP) Observations: Cosmological Interpretation, 
Astrophys. J. Suppl. {\bf 180}, 330 (2009).	
\bibitem {ref14}  
G. Hinshaw, et al., Five-Year Wilkinson Microwave Anisotropy Probe Observations: Data Processing, Sky Maps, and Basic Results, 
Astrophys. J. Suppl. {\bf 180}, 225 (2009). 
\bibitem{ref15} 
A.G. Riess, et al., Type Ia supernova discoveries at $z>1$ from the Hubble space telescope: Evidence for past Deceleration 
and constraints on dark energy evolution, {\it Astrophys. J.} {\bf 607}, (2004) 665.
\bibitem{ref16} 
A. G. Riess, {\it et al.},  New Hubble space telescope discoveries of type Ia supernovae at $z > 1$: narrowing constraints on 
the early behavior of dark energy, Astrophys. J. {\bf 659}, 98 (2007).	
\bibitem {ref17}
P. Garg, R. Zia, and A. Pradhan, Transit cosmological models in FRW universe under the two-fluid scenario, Int. J. Geom. Methods Mod. Phys. 
{\bf 16}, 1950007 (2019).

\bibitem {ref18} 
L. Parker, Particle creation in expanding universes, Phys. Rev. Lett. {\bf 21}, 562 (1968).
\bibitem {ref19} 
V.N. Lukash and A.A. Starobinsky, Isotropization of cosmological expansion due to particle production, 
{\it Zhurnal Ehksperimental'noj i Teoreticheskoj Fiziki} {\bf 66}, 1515 (1974).
\bibitem {ref20} 
J.K. Singh, R. Nagpal, and S.K.J. Pacif, Statefinder diagnostic for modified Chaplygin gas cosmology in $f(R, T)$ gravity with particle creation, 
Int. J. Geom. Methods Mod. Phys. {\bf 15}, 1850049 (2018).

\bibitem {ref21}
W. Zimdahl, Cosmological particle production, causal thermodynamics, and inflationary expansion, Phys. Rev. D {\bf 61}, 083511 (2000).
\bibitem {ref22}
 R.C. Nunes, Connecting inflation with late cosmic acceleration by particle production, Int. J. Mod. Phys. D {\bf 25}, 1650067 (2016).
 \bibitem {ref23}
 G. Steigman, R.C. Santos, and J.A.S. Lima, An accelerating cosmology without dark energy, J. Cosmol. Astropart. Phys. {\bf 06}, 033 (2009).
\bibitem {ref24}
J.A.S. Lima, J.F. Jesus, and F.A. Oliveira, CDM Accelerating cosmology as an Alternative to $\Lambda{CDM}$ model, 
J. Cosmol. Astropart. Phys. {\bf 11}, 0911 (2010). 
\bibitem {ref25} 
J.A.S. Lima, L.L. Graef, D. Pavon, and S. Basilakos, Cosmic acceleration without dark energy: background tests and thermodynamic analysis, 
J. Cosmol. Astropart. Phys. {\bf 10}, 042 (2014).
\bibitem {ref26} 
R.C. Nunes and S. Pan, New observational constraints on f (T) gravity from cosmic chronometers, Mon. Not. Roy. Astron. Soc. {\bf 2016}, 001 (2016).
\bibitem {ref27} 
J. de Haro and S. Pan, Gravitationally induced adiabatic particle production: From Big Bang to de Sitter, Class. Quant. Grav. {\bf 33}, 165007 (2016).
\bibitem {ref28}
 S. Pan, J. de Haro, and A. Paliathanasis, Two-fluid solutions of particle-creation cosmologies, Eur. Phys. J. C {\bf 79}, 115 (2019).
\bibitem {ref28a}
M.R. Setare and A.A. Saharian. Particle creation in an oscillating spherical cavity, Mod. Phys. Lett. A {\bf 16}, 1269 (2001).
\bibitem {ref28b}
M. Mohsenzadeh, E. Yusofi, and M.R. Tanhayi. Particle creation with excited de Sitter modes, Can. J. Phys. {\bf 93}, 1466 (2015).
\bibitem {ref28c}
M.R. Setare and M.J.S. Houndjo. Particle creation in flat Friedmann-Robertson-Walker (FRW) universe in the framework of $f(T)$ gravity,
Can. J. Phys. {\bf 91}, 168 (2013).
 \bibitem {ref28d}
C.P. Singh and A. Beesha, Early universe cosmology with particle creation: kinematics tests, Astrophys. Space Sci. {\bf 336}, 469 (2011).
\bibitem {ref28e}
C.P. Singh and A. Beesham, Particle creation in higher dimensional space-time with variable $G$ and $\Lambda$, Int. J. Theor. Phys. {\bf 51}, 3951 (2012).
\bibitem {ref28f} 
C.P. Singh, FRW models with particle creation in Brans-Dicke theory, Astrophys. Space Sci. {\bf 338}, 411 (2012).
\bibitem {ref28g}
C.P. Singh, Viscous FRW models with Particle creation in early Universe, Mod. Phys. Lett. A {\bf 27}, 1250070 (2012).
\bibitem {ref28h}
V. Singh and C.P. Singh, Friedmann cosmology with matter creation in modified $f(R, T)$ gravity, Int. J. Theor. Phys. {\bf 55}, 1257 (2016).
\bibitem {ref28i}
R.C. Nunes, Connecting inflation with late cosmic acceleration by particle production, Int. J. Mod. Phys. D {\bf 25}, 1650067 (2016).
\bibitem {ref28j}
R.C. Nunes and S. Pan, Cosmological consequences of an adiabatic matter creation process, Mon. Not. Roy. Astron. Soc. {\bf 459}, 673 (2016).
\bibitem {ref28k}
K. Desikan, Cosmological models with bulk viscosity in the presence of particle creation, Gen. Relativ. Grav. {\bf 29}, 435 (1997).
\bibitem {ref28l}
A. Beesham, Bulk viscosity and particle creation in Brans-Dicke theory, Aust. J. Phys. {\bf 52}, 1039 (1999).
 
\bibitem {ref29} 
I. Prigogine, J. Geheniau, E. Gunzig, and P. Nardone, Thermodynamics of cosmological matter creation, Proc. Natl. Acad. Sci. {\bf 85}, 7428 (1988).
\bibitem {ref30} 
 I. Prigogine, et al., Thermodynamics and cosmology, Gen. Relativ. Grav. {\bf 27}, 767 (1989).
\bibitem {ref31} 
E. Schrödinger,  The proper vibrations of the expanding universe, Physica {\bf 6}, 899 (1939).
\bibitem {ref32}
N. D. Birrell and P.C.W. Davies, Quantum fields in curved space, Cambridge University Press (1982).
\bibitem {ref33}
L.H. Ford, D3: Quantum field theory in  curved spacetime,  Gen. Relat. Gravit. 490 (2002).
\bibitem {ref34} 
Yu.V. Pavlov, Space-time description of scalar particle creation by a homogeneous isotropic gravitational field, 
Grav, Cosmol. {\bf 14}, 314 (2008).
 \bibitem {ref35}
 A. Dixit, P. Garg, and A. Pradhan, Particle creation in FLRW higher dimensional universe with gravitational and cosmological constants, 
  Can. J. Phys. Accepted on 24 Feb. (2021).
 \bibitem {ref36}
V.B. Johri and K. Desikan,  An Extended Class of FRW Models with Creation of Particles out of Gravitational Energy, 
Astro. Lett. Comm. {\bf 33}, 287 (1996).
\bibitem {ref37}
J.A.S. Lima and J.S. Alcaniz, Flat FRW cosmologies with adiabatic matter creation: Kinematic tests, Astron. Astrophys. {\bf 348}, 1 (1999).
\bibitem {ref38}
J.S. Alcaniz and J.A.S. Lima, Closed and open FRW cosmologies with matter creation: kinematic tests, 
Astron. Astrophys. {\bf 349}, 729 (1999).
\bibitem {ref39}
W. Zimdahl, et al., Cosmic antifriction and accelerated expansion, Phys. Rev. D {\bf64}, 063501 (2001).

\bibitem {ref40}
Yuan Qiang, Tong-Jiezhang, and Yi. Ze-Long, Constraint on cosmological model with matter creation using complementary astronomical 
observations, Astrophys. Space Sci. {\bf 311}, 407 (2007). 
\bibitem {ref41}
E. Aydiner, Chaotic universe model, Scientific Reports {\bf 8}, 721 (2018). 
\bibitem {ref42} 
S. Pan, B.K. Pal, and S. Pramanik, Gravitationally influenced particle creation models and late-time cosmic acceleration, 
Int. J. Geom. Methods Mod. Phys. {\bf 15}, 1850042 (2018).
\bibitem {ref43}
V. Sahni and S. Habib, Does inflationary particle production suggest $\Omega_{m}<1 ?$, Phys. Rev. Lett. {\bf 81},
1766 (1998), arXiv:hep-th/9808204. 
\bibitem {ref44}
 T. Piran and S. Weinberg, et al., Physics in Higher Dimensions, World Scientific, {\bf 2} (1986).
 \bibitem {ref45}
 T. Kaluza, On the unification problem in Physics, Int. J. Mod. Phys. D {\bf 27}, 1870001 (2018) (translation), 
 Sitzungsber. Preuss. Akad. Wiss. Berlin (Math. Phys.) 966 (1921) (original).
 \bibitem {ref46}
O. Klein, Quantum theory and five dimensional theory of Relativity, Z. Phys. {\bf 37}, 895 (1926).
 \bibitem {ref47}
E. Witten, Fermion numbers in Kaluza-Klein theory, Phys. Lett. B {\bf 144}, 351 (1984).
\bibitem {ref48}
 T. Appelquist, A. Chodos, and P. G. O. Freund, Modern Kaluza-Klein Theories (Frontiers in Physics), Addison-Wesley (1987)
 \bibitem {ref49}
 J. M. Overduin, P. S. Wesson, Kaluza-klein gravity, Phys. Rep. {\bf 283}, 303 (1997).
\bibitem {ref50}
S. Chatterjee and B. Bhui, Homogeneous cosmological model in higher dimension, Mon. Not. R. Astron. Soc. {\bf 247}, 57 (1990).
\bibitem {ref51}
O. Sevinc and E. Aydiner, Particle Creation in Friedmann-Robertson-Walker Universe { Grav. $\&$ Cosmo.} {\bf 25} (2019) 397.

 \bibitem {ref52}
 A. Saha and S. Ghose, Interacting Tsallis holographic dark energy in higher dimensional Cosmology, Astrophys. Space Sci. {\bf 365}, 98 (2020) 98.
 
\bibitem {ref53}
 G.P. Singh and S. Kotambkar, Higher-Dimensional Dissipative Cosmology with Varying G and $\Lambda$, Gravit. Cosmol. {\bf 9}, 206 (2003).
\bibitem {ref54}
G.P. Singh, S. Kotambkar, and A. Pradhan, Higher dimensional cosmological model in Lyra Geometry:
Revisited, Int. J. Mod. Phys. D {\bf 12}, 853 (2003).
 \bibitem {ref55}
G.S. Khadekar, V. Kamdi, A. Pradhan, and S. Otarod, Five dimensional universe model with variable
cosmological term $\Lambda$ and a big bounce, Astrophys. Space Sci. {\bf 310}, 141 (2007).
\bibitem {ref56}
T. Harko and M.K. Mak, Particle creation in cosmological models with varying gravitational and cosmological
constant, Gen. Rel. Grav. {\bf 31}, 6 (1999).
\bibitem {ref57}
S. Nojiri and S.D. Odintsov, Phys. Dark Energy {\bf 30} 100695 (2020).
\bibitem {ref58}
R. Caldwell, A Phantom Menace? Cosmological consequences of a dark energy component with super-negative equation of state, 
Phys. Lett. B {\bf 545}, 23 (2002).
\bibitem {ref59}
 M. Shen, L. Zhao, Oscillating quintom model with time periodic varying deceleration parameter, Chin. Phys. Lett. {\bf 31}, 010401 (2014).
 \bibitem {ref59a}
 P.K. Sahoo, S.K. Tripathi, and P. Sahoo, A periodic varying deceleration parameter in $f(R, T)$ gravity, Mod. Phys. Lett. A {\bf 33}, 
 1850193 (2018), arXiv:1710.09719[gr-qc].
 
\bibitem {ref60}
S. Dodelson, M. Kaplinghat, and E. Stewart, Solving the coincidence problem: Tracking oscillating energy,  Phys. Rev. Lett. {\bf 85}, 5276 (2000).
\bibitem {ref61}
 S. Nojiri and S.D. Odintsov, The oscillating dark energy: future singularity and coincidence problem, Phys. Lett. B {\bf 637}, 139 (2006).

\bibitem {ref62}
J.A.S. Lima, M.O. Calvao, and I. Waga, Frontier Physics, Essay in Honor of Jayme Tiomno (World Scientific, Singapore, (1990).

\bibitem {ref63}
 H. Yu,  B. Ratra, and F. Wang, Hubble parameter and Baryon Acoustic Oscillation measurement constraints on the Hubble constant, 
 the deviation from the spatially flat $\Lambda C D M$ model, the deceleration acceleration transition redshift, and spatial curvature, 
 Astrophys. Journ. {\bf 856}, 3 (2018).
\bibitem {ref64}
 H. Amirhashchi and S. Amirhashchi, Constraining Bianchi Type I universe with Type Ia supernova and $H(z)$ data, 
 Phys. Dark Univ. {\bf 29}, 100557 (2020), arXiv: 1802.04251[ astro-ph.CO ] .
\bibitem {ref65}
R. Rebelo, Nucl. Phys. B, Proc. Suppl. {\bf 114}, 3 (2003).
\bibitem {ref66}
 L.V.E. Koopmans, et al., The Hubble Constant from the Gravitational Lens B1608$\pm$656,  Astrophys. J. {\bf 599}, 70 (2003). 

\bibitem {ref67}
S.M. Carroll and M. Hoffman, Can the dark energy equation-of-state parameter $w$ be less than $-1 ?$ Phys. Rev. D {\bf68}, 023509 (2003).
\bibitem {ref68}
G.K. Goswami, A. Pradhan, and A. Beesham, A dark energy quintessence model of the universe,
Mod. Phys. Lett. A {\bf 35}, 2050002 (2020).
\bibitem {ref69}
G.K. Goswami, M. Mishra, A.K. Yadav, and A Pradhan, Two-fluid scenario in Bianchi type-I
universe, Mod. Phys. Lett. A {\bf 35}, 2050086 (2020).
\bibitem {ref70}
G.K. Goswami, R.N. Dewangan, and A.K. Yadav, Anisotropic universe with magnetized dark energy,
Astrophys. Space Sci. {\bf 361}, 119 (2016).
\bibitem {ref71}
J.J. Condon and A.M. Matthews, $\Lambda$CDM cosmology for astronomers, Public. Astronom. Society Pacific
{\bf 130}, 073001 (2018), arXiv:1804.10047[astro-ph.CO].
\bibitem {ref72}
S. Nojiri, S.D. Odintsov, and S. Tsujikawa, Properties of singularities in (phantom) dark energy universe, Phys. Rev. D {\bf 71}, 063004 (2005).

\end{thebibliography}
\end{document}